\documentclass[9pt]{article}
\usepackage{float}
\usepackage[symbol]{footmisc}
\usepackage{algorithm}
\usepackage{algpseudocode}
\usepackage[superscript,sort]{cite}
\usepackage{array}
\usepackage[pdftex]{graphicx}
\usepackage[usenames,pdftex]{color}
\definecolor{Red}{rgb}{1.0,0.0,0.0}
\usepackage{amsmath,amssymb}

\usepackage{amsthm,amscd,amsxtra,amsfonts,amsmath,amssymb,multirow}
\usepackage{wrapfig}
\usepackage[footnotesize]{caption}
\usepackage[tiny,compact]{titlesec}
\usepackage{ctable}

\usepackage{xcolor,colortbl}
\usepackage{subfigure}
\usepackage{tikz}
\usetikzlibrary{shapes.geometric, arrows}
\usepackage[T1]{fontenc}

\titlespacing*{\section}{0pt}{*0}{*0}
\titlespacing*{\subsection}{0pt}{*0}{*0}
\titlespacing*{\subsubsection}{0pt}{*0}{*0}
\titlespacing{\paragraph}{0pt}{*0}{*1}

\definecolor{MyPurple}{rgb}{1,0,1}

\newcommand{\beq}[1]{\begin{equation} \label{#1}}
\newcommand{\eeq}{\end{equation}}
\newcommand{\barray}{\begin{array}{ll}}
\newcommand{\earray}{\end{array}}

\begin{document}
\pagenumbering{roman}

 \renewcommand{\thepage}{{\arabic{page}}}

\title{Multiscale Gaussian network model (mGNM) and multiscale anisotropic network model (mANM)
}

\author{
Kelin Xia$^1$, Kristopher Opron$^2$ and
Guo-Wei Wei$^{3}$ \footnote{ On leave from  Department of Mathematics, Michigan State University}~\footnote{ Address correspondences  to Guo-Wei Wei. E-mail:wei@math.msu.edu}\\
$^1$ Department of Mathematics \\
Michigan State University, East Lansing, MI 48824, USA\\
$^2$  Department of Biochemistry and Molecular Biology\\
Michigan State University, East Lansing, MI 48824, USA \\
$^3$ Mathematical Biosciences Institute\\
The Ohio State University,
Columbus, Ohio 43210, USA
}

\date{\today}
\maketitle

\begin{abstract}

Gaussian network model (GNM) and anisotropic network model (ANM) are some of the most popular methods for the study of protein flexibility and related functions. In this work, we propose   generalized GNM (gGNM) and ANM methods and  show that the GNM Kirchhoff matrix can be built  from the ideal low-pass filter, which is a special case of a wide class of correlation functions underpinning the linear scaling flexibility-rigidity index (FRI) method. Based on the mathematical structure of  correlation functions,  we propose a unified framework to construct generalized Kirchhoff matrices whose matrix inverse leads to gGNMs, whereas, the direct inverse of its diagonal elements gives rise to FRI method. With this connection, we further introduce two multiscale elastic network models, namely, multiscale GNM (mGNM) and multiscale ANM (mANM), which are able to incorporate different scales into the generalized Kirchkoff matrices or generalized Hessian matrices. We validate our new multiscale methods with extensive numerical experiments. 
We illustrate that gGNMs outperform the original GNM method in the B-factor prediction of a set of 364 proteins. We demonstrate that for a given correlation function, FRI and gGNM methods provide essentially identical B-factor predictions when the scale value in the correlation function is sufficiently large. More importantly, we reveal intrinsic multiscale behavior in  protein structures. The proposed mGNM and mANM are able  to  capture  this multiscale behavior and thus give rise  to a significant improvement of more than 11\%  in B-factor predictions over the original GNM and ANM methods. We further demonstrate benefit of our mGNM  in the B-factor predictions on many proteins that fail the original GNM method. We show that the present mGNM can also be used to analyze protein domain separations.  Finally, we showcase the ability of our mANM  for the simulation of protein collective motions.

\end{abstract}


%
%

\newpage

 \section{Introduction}\label{sec:Intro}


Under physiological conditions, proteins undergo everlasting motions, ranging from atomic thermal fluctuation, side-chain rotation, residue swiveling, to domain swirling. protein motion strongly correlates with protein functions, including  molecular docking \cite{Fischer:2014}, drug binding \cite{Alvarez-Garcia:2014},  allosteric signaling \cite{ZBu:2011}, self assembly \cite{Marsh:2014} and enzyme catalysis \cite{Fraser:2009}. The range of protein motions in a cellular environment depends on the structure's local flexibility, an intrinsic property of a given protein structure. Protein flexibility is reflected by the Debye-Waller factor (B-factor), i.e., the atomic mean-square displacement, obtained in  structure determination by x-ray crystallography, NMR, or single-molecule force experiments \cite{Dudko:2006}. However, the B-factor cannot absolutely quantify flexibility: it also depends the crystal environment, solvent type, data collection condition and structural refinement procedure \cite{Kondrashov:2007,Hinsen:2008}.

The flexibility of a biomolecule can be  assessed by molecular dynamics (MD) \cite{McCammon:1977},   normal mode analysis (NMA)  \cite{Go:1983,Tasumi:1982,Brooks:1983,Levitt:1985}, graph theory \cite{Jacobs:2001} and elastic network models (ENMs) \cite{Bahar:1997,Bahar:1998,Atilgan:2001,Hinsen:1998,Tama:2001,LiGH:2002}, including Gaussian network model (GNM) \cite{Bahar:1997,Bahar:1998} and  anisotropic network model (ANM) \cite{Atilgan:2001}. NMA  can be regarded as time-independent molecular dynamics (MD) \cite{JKPark:2013} and diagonalizes the MD potential to obtain a set of eigenvalues and eigenvectors, where first  few  eigenvectors predict the collective and global motions,   which are  potentially relevant to   biomolecular functionality. NMA with only the elasticity potential, which is a leading term in the MD potential, was introduced by Tirion \cite{Tirion:1996}, and was extended to the network setting in ANM \cite{Atilgan:2001}.  Here network refers to the connectivity between particles regardless of their chemical bonds \cite{Flory:1976}. Yang et al. \cite{LWYang:2008} demonstrated that due to its network setting, GNM is about one order more efficient than most other flexibility approaches in computational complexity. In terms of B-factor prediction, GNM is typically more accurate than ANM \cite{JKPark:2013, Opron:2014}. Therefore, GNM has been widely used in the study of biomolecular structure, dynamics and functions \cite{JMa:2005,LWYang:2008,Skjaven:2009,QCui:2010}. It has demonstrated its utilities in stability analysis \cite{Livesay:2004} ,  docking simulation  \cite{Gerek:2010},  viral capsids \cite{Rader:2005,Tama:2005} and domain motions of  hemoglobin \cite{CXu:2003}, F1 ATPase \cite{WZheng:2003,QCui:2004}, chaperonin GroEL \cite{Keskin:2002,WZheng:2007} and the ribosome \cite{Tama:2003,YWang:2004}.

In traditional elasticity network models, i.e., ENM, GNM and ANM, the connectivity is determined by a fixed cutoff distance. All atoms within the cutoff distance are treated equally with no consideration given to effects which scale with distance. In this manner, the behavior of these methods typically depends on the cutoff distance used.  Many modified models are proposed  to improve the hard cutoff distance practice by the incorporation of the distance information \cite{LWYang:2008}.    Hinsen  has changed the spring constant to a distance-dependent Gaussian function \cite{Hinsen:1998}. Riccardi et al.  have used an inverse 6th power function of distance as the spring coefficient in their elastic network model \cite{Riccardi:2009}. A parameter-free ANM has been introduced by  using the inverse 2nd power square distance for spring constant  \cite{LYang:2009}.

A common feature of all the aforementioned  approaches  is that, they all depend on the mode decomposition of the potential matrix, which typically has the computational complexity of $O(N^3)$, where $N$ is the  order the potential matrix. To  bypass the matrix diagonalization, researchers explore the flexibility  properties using the local packing density. Many elegant methods and algorithms have been proposed, including the local density model (LDM) by Halle \cite{Halle:2002}, the local contact model (LCM) by Zhang et al \cite{FLZhang:2002},  weighted contact number (WCN) by \cite{CPLin:2008}, and others \cite{CPLin:2008,SWHuang:2008,DWLi:2009}.

Recently, we have proposed a few mode-decomposition free methods for flexibility  analysis, namely,  molecular nonlinear dynamics \cite{KLXia:2013b},  stochastic dynamics \cite{KLXia:2013f} and flexibility-rigidity index  (FRI)  \cite{KLXia:2013d,Opron:2014}. Among them, the FRI method is of $O(N^2)$ in computational complexity and has been accelerated to $O(N)$ by using the cell lists algorithm without  loss of accuracy  \cite{Opron:2014}. The essential idea of the FRI method is to evaluate the rigidity index or the compactness of the biomolecular (network) packing by the total correlation, a function of inter-atomic distance. Then the flexibility index is defined as the inverse of the rigidity index.

Although the original motivation for FRI comes from the design of continuum elasticity with atomic rigidity (CEWAR), FRI shares a similarity with the LDM, LCM and WCN. To be more specific, all of these methods make use of local packing information and are free from matrix diagonalization. However, significant distinction exists between our FRI methods and other local density based methods and it can be summarized as the following. Firstly, other than the discrete flexibility index and discrete rigidity index, our FRI methods delivers a continuous flexibility function and a continuous rigidity function  \cite{KLXia:2013d,Opron:2014}. The continuous rigidity function, which can be regarded as the density distribution function (density estimator) of a biomolecule, plays many important roles beyond the scope of flexibility study \cite{KLXia:2014c}. For instance, it can be used to generate biomolecular surface representations \cite{KLXia:2015b,KLXia:2015c}, which reduce to the Gaussian surface if an appropriate kernel is used. In fact, rigidity function can be applied to decipher the atomic information from the experimental electron density data \cite{MTopf:2008,Wriggers:1999,KLXia:2015b}. Secondly, protein multiscale collective motions  can be captured by using multiple kernels in our FRI method, called multiscale FRI or multikernel FRI (mFRI) \cite{Opron:2015a}. This approach significantly improve the accuracy of FRI B-factor predictions. Thirdly, we proposed an anisotropic FRI  (aFRI) model for the evaluation of biomolecular global motions. Different from traditional normal mode analysis or ANM, our aFRI allows adaptive representations, ranging from a completely global description to a completely local representation \cite{Opron:2014}. 

The objective of the present work is twofold. First, we  propose a unified framework to construct   generalized  GNMs (gGNMs).   We reveal that the GNM Kirchhoff matrix can be constructed from the ideal low-pass filter (ILF), which is the limiting case of  admissible FRI correlation  functions. We demonstrate that FRI and gGNM are asymptotically equivalent when the cutoff value in the Kirchhoff matrix or the scale value in the correlation function is sufficiently large. This finding  paves the way for understanding the connection between the GNM and FRI methods. To clarify this connection, we introduce a generalized Kirchhoff matrix to provide a unified starting point for the gGNM and FRI methods, which elucidates on the similarity and difference between gGNM and FRI. Based on this new  understanding of the gGNM working principle, we propose infinitely many correlation function based gGNMs. We show that gGNM outperforms the original GNM for the B-factor prediction of a set of 364 proteins. Both gGNM and FRI deliver almost identical results when the scale parameter is sufficiently large. Our approach sheds light on the construction of efficient gGNMs.
Additionally, we propose two new methods, multiscale GNM (mGNM) and multiscale ANM (mANM), to account for the multiscale features of biomolecules.  Most biomolecules, particularly large macromolecules and protein complexes, have multiple  characteristic length scales ranging from covalent bond, residue, secondary structure and domain dimensions, to protein sizes.  Even for small molecules, due to the influence of crystal structure, multiscale effects play a significant role in atomic thermal fluctuations.  Consequently, GNM and ANM, which are typically parametrized at a single cutoff distance, often do not work well  in characterizing the flexibility of molecules involving multiscale behaviors. Our essential idea is to generalize original GNM and ANM into a multikernel setting so that each kernel can be parametrized at a given characteristic length. This generalization is achieved through the use of an FRI assessment, which predicts the involvement of different scales, followed by an appropriate constructions of multikernel GNM or multikernel ANM.
This approach works because   for a diagonally dominant matrix,  the direct inverse of the diagonal element is essentially equivalent to the diagonal element of the inverse matrix.  We demonstrate that the proposed mGNM and mANM are able to successfully capture the multiscale properties of the protein and significantly improve the accuracy in protein flexibility prediction.

The rest of this paper is organized as the follows. Section \ref{sec:method} is devoted to methods and algorithms. We first propose a concise formulation of gGNMs using FRI correlation functions in Section \ref{sec:GNM}. We show that there are infinitely many new gGNMs that reduce to the original GNM at appropriate limits of their parameters.  To establish notation,  we further present a brief review of our mFRI formalism in Section \ref{sec:mFRI}.
Based on the connection between  FRI and GNM, we propose mGNM in Section \ref{sec:mGNM}. 
Specifically, parameters learned from mFRI are used to construct the multiscale Kirchhoff matrix in mGNM.
We discuss two types of  realizations of mGNMs.
As an extension of our mGNM, the mANM method is introduced in Section \ref{sec:mANM}.
We validate the proposed gGNM, mGNM and mANM by extensive numerical experiments in Section \ref{sec:result}. We illustrate that the intrinsic multiscale properties of  biomolecules are successfully captured in our mGNM and mANM. Finally, in Section \ref{sec:application}, we demonstrate  the utility of mGNM and mANM for protein flexibility analysis, protein domain separation and collective motion study. The present work offers a new strategy for the design and construction of  accurate, efficient and robust  methods for biomolecular flexibility analysis. This paper ends with a conclusion.

\section{Methods and algorithms}\label{sec:method}

\subsection{Generalized Gaussian network models (gGNMs)}\label{sec:GNM}
To establish notation and facilitate new development, let us present a brief review of the GNM and FRI methods. Consider an  $N$-particle coarse-grained representation of a biomolecule. We denote $\{ {\bf r}_{i}| {\bf r}_{i}\in \mathbb{R}^{3}, i=1,2,\cdots, N\}$ the coordinates of these particles and $r_{ij}=\|{\bf r}_i-{\bf r}_j\|$ the Euclidean space distance between  $i$th   and $j$th particles. In a nutshell, the GNM prediction of the $i$th B-factor of the biomolecule can be expressed as \cite{Bahar:1997,Bahar:1998}
\begin{eqnarray}\label{eqn:GNM}
B_i^{\rm GNM}=a \left(\Gamma^{-1} \right)_{ii}, \forall i=1,2,\cdots, N,
\end{eqnarray}
where $a$ is a fitting parameter that can be related to  the thermal energy and $\left(\Gamma^{-1} \right)_{ii}$ is the $i$th diagonal element of the matrix inverse of the Kirchhoff matrix,
\begin{eqnarray}\label{eqn:Kirchhoff}
\Gamma_{ij}  = \begin{cases}\begin{array}{ll}
       -1, &i\neq j ~{\rm and} ~r_{ij} \leq r_c \\
       0,  &    i\neq j ~{\rm and} ~r_{ij}  > r_c  \\
        -\sum_{j, j\neq i}^N\Gamma_{ij},  & i=j
							\end{array}
       \end{cases},
\end{eqnarray}
where $r_c$ is a cutoff distance.  The GNM theory evaluates the  matrix inverse by  $\left(\Gamma^{-1} \right)_{ii}=\sum_{k=2}^N  \lambda_k^{-1}\left[{\bf u}_k {\bf u}_k^T \right]_{ii}$, where $T$ is the transpose and $\lambda_k$ and ${\bf u}_k$ are the $k$th eigenvalue and eigenvector of $\Gamma$, respectively. The summation omits the first eignmode whose eigenvalue is zero.

The FRI   prediction of the $i$th B-factor of the biomolecule can be given by \cite{KLXia:2013d,Opron:2014}
\begin{eqnarray}\label{eqn:FRI}
B_i^{\rm FRI}=a \frac{1}{\sum_{j,j\neq i}^N w_j\Phi(r_{ij};\eta)} + b, \forall i=1,2,\cdots, N,
\end{eqnarray}
where $a$ and $b$ are fitting parameters,   $f_i=\frac{1}{\sum_{j,j\neq i}^N w_j\Phi(r_{ij};\eta)}$ is the $i$th flexibility   index and $\mu_i=\sum_{j,j\neq i}^N w_j\Phi(r_{ij};\eta)$ is the $i$th rigidity index.
Here, $w_j$ is an atomic number depended weight function that can be set to $w_j=1$ for a C$_{\alpha}$ network, and $\Phi(r_{ij};\eta)$ is  a    real-valued monotonically decreasing correlation function satisfying the following admissibility conditions
\begin{eqnarray}\label{eq:couple_matrix1-1}
\Phi( r_{ij};\eta)&=&1 \quad {\rm as }\quad  r_{ij}   \rightarrow 0\\ \label{eq:couple_matrix1-2}
\Phi( r_{ij};\eta)&=&0 \quad {\rm as }\quad  r_{ij}   \rightarrow\infty,
\end{eqnarray}
where $\eta$ is a scale parameter. Delta sequences of the positive type  \cite{GWei:2000} are good choices. Many radial basis functions are also admissible   \cite{KLXia:2013d,Opron:2014}.  Commonly used FRI correlation functions include the  generalized exponential  functions
\begin{eqnarray}\label{eq:couple_matrix1}
\Phi( r_{ij};\eta, \kappa) =    e^{-\left( r_{ij} /\eta \right)^\kappa},    \quad \kappa >0,
\end{eqnarray}
and  generalized Lorentz functions
\begin{eqnarray}\label{eq:couple_matrix2}
 \Phi( r_{ij};\eta, \upsilon) =  \frac{1}{1+ \left( r_{ij} /\eta\right)^{\upsilon}},  \quad  \upsilon >0.
 \end{eqnarray}
A major advantage of the FRI method is that it does not resort to  mode decomposition and its computational complexity can be reduced to $O(N)$ by means of the cell lists algorithm used in our fast FRI (fFRI) \cite{Opron:2014}. In contrast, the mode decomposition of NMA and GNM has the computational complexity of $O(N^3)$.

 \begin{figure}[]
\begin{center}
\begin{tabular}{c}
\includegraphics[width=0.8\textwidth]{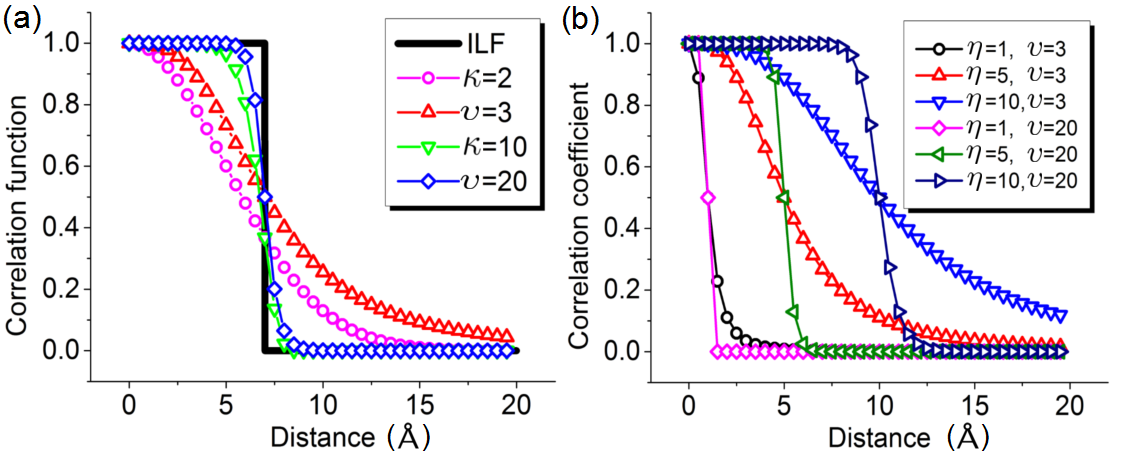}
\end{tabular}
\end{center}
\caption{Illustration of admissible correlation functions.
(a) Correlation  functions approach the ILF as $\kappa\rightarrow \infty$ or  $\upsilon\rightarrow \infty$ at $\eta=7$\AA.
(b) Effects of varying scale value $\eta$. Local correlation is obtained with large $\upsilon$ and small $\eta$ values. Whereas, nonlocal correlation is generated by small $\upsilon$ and large $\eta$ values.
 }
\label{functions}
\end{figure}
To further explore the theoretical foundation of GNM, let us examine the parameter limits of   generalized exponential functions (\ref{eq:couple_matrix1}) and generalized Lorentz functions (\ref{eq:couple_matrix2})
\begin{eqnarray}\label{eq:Asmpt1}
e^{-\left( r_{ij} /\eta\right)^\kappa} \rightarrow \Phi( r_{ij};r_c)  & {\rm as } & \kappa\rightarrow\infty\\ \label{eq:Asmpt2}
\frac{1}{1+ \left( r_{ij} /\eta\right)^{\upsilon}} \rightarrow \Phi( r_{ij};r_c) &  {\rm as } &\upsilon\rightarrow\infty,
\end{eqnarray}
where $r_c=\eta$ and $\Phi( r_{ij};r_c) $ is  the ideal low-pass  filter (ILF) used in the GNM Kirchhoff matrix
\begin{eqnarray}\label{eqn:IdealLF}
\Phi( r_{ij};r_c)  = \begin{cases}\begin{array}{ll}
       1, &  r_{ij} \leq r_c \\
       0,  & r_{ij}  > r_c  \\
			\end{array}
       \end{cases}.
\end{eqnarray}
Relations (\ref{eq:Asmpt1}) and (\ref{eq:Asmpt2}) unequivocally connect FRI correlation functions to the GNM Kirchhoff matrix. It is important to examine whether the ILF is still an FRI correlation function.  Mathematically, the ILF is a special real-valued monotonically decreasing correlation function and also satisfies admissibility conditions  (\ref{eq:couple_matrix1-1}) and (\ref{eq:couple_matrix1-2}). In fact,  all FRI correlation functions are  low-pass filters as well. Therefore, both   GNM and   FRI admit low-pass filters in their constructions. Indeed, GNM is very special in the sense that there is only one unique ILF, while, there are infinitely many other low-pass filters. Figure \ref{functions} illustrates the behavior and relation of the above low-pass filters or  correlation functions. Clearly, the ILF is completely localized for any given cutoff value. In general,   generalized exponential function and generalized Lorentz function are delocalized and the former decays faster than the latter for a given power. The combination of a low power value and a large scale gives rise to nonlocal correlations.  Our earlier test indicates that $\upsilon=3$ and $\eta=3$\AA~ provides a good flexibility analysis for   a set of 364 proteins \cite{Opron:2014}.


To further   bring to light the mathematical foundation of the GNM and FRI methods,  we  consider a generalized Kirchhoff matrix  \cite{KLXia:2013f, KLXia:2013b}
\begin{eqnarray}\label{eqn:MKirchhoff}
\Gamma_{ij}(\Phi )  = \begin{cases}\begin{array}{ll}
       - \Phi( r_{ij};\eta), &i\neq j  \\
        -\sum_{j, j\neq i}^N\Gamma_{ij}(\Phi),  & i=j
							\end{array}
       \end{cases},
\end{eqnarray}
where $\Phi( r_{ij};\eta)$ is an admissible FRI correlation function.  The generalized Kirchhoff matrix includes the Kirchhoff matrix as a special  case. It is important to note  that  each diagonal element is an  FRI rigidity index: $\mu_i=\Gamma_{ii}(\Phi )$. Therefore,  the generalized Kirchhoff matrix   provides a unified starting point for both the FRI and gGNM methods. However, the striking difference between the gGNM and FRI methods is that  to predict B-factors, the gGNM seeks a matrix inverse of the Kirchhoff matrix (\ref{eqn:Kirchhoff}), whereas, the FRI takes the direct inverse of the diagonal   elements of the generalized  Kirchhoff matrix (\ref{eqn:MKirchhoff}).

\subsection{Multiscale flexibility-rigidity index (mFRI)}\label{sec:mFRI}
Due to the widely existed multiscale in biomolecules, especially large macromolecules and protein complexes, the multiscale FRI (mFRI) method is proposed to better capture nonlocal collective mentions \cite{Opron:2015a}. In this approach, two or three correlation kernels that are parametrized at multiple scales are employed simultaneously to characterize protein multiscale properties. The flexibility index can be expressed as,
\begin{eqnarray}\label{eq:flexibility3}
 f^{n}_i & = & \frac{1}{\sum_{j,j\neq i}^N w^{n}_{j} \Phi^{n}( r_{ij};\eta^{n}  )}, \forall i=1,2,\cdots, N,
 \end{eqnarray}
where  $w^{n}_{j}$, $\Phi^{n}( r_{ij};\eta^{n} ) $ and $\eta^{n} $ are the corresponding quantities associated with the $n$th kernel.
The essence of the mFRI is to minimize of the following form
\begin{eqnarray}\label{eq:fit_mFRI}
{\rm Min}_{a^{n},b} \left\{ \sum_i \left| \sum_{n}a^n f^{n}_i + b-B^e_i\right|^2\right\},
\end{eqnarray}
where  $B^e_i $ are the experimental B-factors for the $i$th particle. 
We choose kernels with various scale parameters and obtain the optimized fitting coefficients.

Specifically, for a coarse-grained network model with only C$_\alpha$ atoms, we can set  $w^n_j=1$ and choose a single type of kernel function parametrized at different scales. The predicted B-factors can be expressed as
\begin{eqnarray}\label{eq:flexibility4}
 B^{\rm mFRI}_i  = b+ \sum_{n=1}\frac{a^n}{\sum_{j,j\neq i}^N  \Phi( r_{ij};\eta^{n} )}, \forall i=1,2,\cdots, N.
 \end{eqnarray}
Unlike the scheme in Eq. (\ref{eq:flexibility3})  where various types of kernels can be chosen, we only select one type of kernels in Eq. (\ref{eq:flexibility4}). In this way, the multiscale nature in biomolecules can be clearly demonstrated. Some commonly used kernel functions include generalized Lorentz kernel,
\begin{eqnarray}\label{eq:couple_matrixn}
 \Phi(\|{\bf r} - {\bf r}_j \|;\eta^n) = \frac{1}{1+ \left( \|{\bf r} - {\bf r}_j \|/\eta^n\right)^{3}},
 \end{eqnarray}
and the generalized exponential kernel,
\begin{eqnarray}\label{eq:couple_matrixn2}
 \Phi(\|{\bf r} - {\bf r}_j \|;\eta^n) = e^{-\frac{\|{\bf r} - {\bf r}_j \|}{\eta^n}}.
 \end{eqnarray}
These kernels define a continuous multiscale rigidity function by using the fitting coefficients from the minimization process as following,
\begin{eqnarray}\label{eq:mrigidity_function}
 \mu({\bf r})  =\sum_{j=1}^N w^{n}_{j} \Phi( \|{\bf r} - {\bf  r}_j \|;\eta^{n} ).
 \end{eqnarray}
This expression can be used to construct new protein surfaces.
Similarly, we can also construct a continuous multiscale flexibility function,
\begin{eqnarray}\label{eq:mflexibility_function}
 f({\bf r})  = b+ \sum_{n=1}\frac{a^n}{\sum_{j=1}^N w^{n}_{j} \Phi( \|{\bf r} - {\bf  r}_j \|;\eta^{n} )}.
 \end{eqnarray}
 One can map this continuous multiscale flexibility function onto a molecular surface to analyze the flexibility of the molecule.

\subsection{Multiscale Gaussian network model (mGNM)}\label{sec:mGNM}

The essential component for our mGNM is to build a multiscale Kirchhoff matrix, which incorporates  various scales instead of a single one. Due to the intrinsic relation between FRI and gGNM discussed in Section \ref{sec:GNM}, we make use of the coefficients approximated from our FRI to construct a multiscale Kirchhoff matrix. In this section, we present two types of algorithms to construct mGNM.

\paragraph{Type-1 mGNM}
First, we assume that the multiscale Kirchhoff matrix takes the form
\begin{equation}\label{type-1mGNM}
\Gamma =\sum_{n}a^n \Gamma^{n},
\end{equation}
where $a^n$ and $\Gamma^{n}=\left(\Gamma_{ij}(\Phi^{n}( r_{ij};\eta^{n}_{j} ))\right)$ are the fitting coefficient and generalized Kirchhoff matrix associated with the $n$th kernel $\Phi^{n}( r_{ij};\eta^{n} ))$ parametrized at an appropriate scale $\eta^{n}$.
We use our  mFRI to evaluate coefficients $\{a^n\}$. Basically, we have multiscale rigidity index $\mu_i=\sum_{n}a^n \Gamma^{n}_{ii}$.  Then, $\{a^n\}$ are determined via the minimization ${\rm Min} \sum_{i}\left| \frac{1}{\mu_i} -{B^e_i}\right|^2$, which is equivalent to
\begin{eqnarray}\label{eq:regression2}
{\rm Min}_{a^{n}} \left\{ \sum_i \left| \sum_{n}a^n \Gamma^{n}_{ii} -\frac{1}{B^e_i}\right|^2\right\},
\end{eqnarray}
assuming that $B^e_i>0$. With the multiscale Kirchhoff matrix given in Eq. (\ref{type-1mGNM}), we carry our routine GNM analysis as described in Eq. (\ref{eqn:GNM}).

\paragraph{Type-2 mGNM}
Another algorithm for constructing mGNM is to make use of fitting coefficients from mFRI directly via  the relation between biomolecular local packing density and its flexibility.
 Basically, we choose several kernels parametrized at various scales and evaluate the best fitting coefficients, i.e., $\{a_n\}$ and $b$, with the experimental B-factors using Equation (\ref{eq:fit_mFRI}). The resulting multiscale flexibility index is then used to construct the generalized Kirchhoff matrix as following
\begin{eqnarray}\label{eq:relation1}
\sum_{n}a^n f^{n}_i + b=\frac{1}{\Gamma_{ii}}, \forall i=1,2,\cdots, N.
 \end{eqnarray}
With the relation $f^{n}_i=\frac{1}{\mu^{n}_i}, \forall i=1,2,\cdots, N$, the above expression can be rewritten as,
 \begin{eqnarray}\label{eq:relation2}
\Gamma_{ii}=\frac{1}{\sum_{n} \frac{a^n}{\mu^{n}_i}+ b}, \forall i=1,2,\cdots, N.
 \end{eqnarray}
Usually, we can use two or three kernels parametrized at different scales. For instance, if we use two kernels, we  can further rewrite the above expression  as,
  \begin{eqnarray}\label{eq:2ker_gnm1}
  \Gamma_{ii}= \frac{\mu^{1}_i\mu^{2}_i}{a^1\mu^{2}_i+a^2\mu^{1}_i +b\mu^{1}_i\mu^{2}_i}, \forall i=1,2,\cdots, N.
 \end{eqnarray}
Now the problem is to determine the non-diagonal terms of our multiscale Kirchhoff matrix. One simple approach is to subdivide either of the two rigidity indices. For example, we can choose to use the rigidity index for the first kernel. Since we have $\mu^{n}_i=\sum_{j,j\neq i}^N w^{n}_{j} \Phi^{n}( r_{ij};\eta^{n} ), ~n=1,2$, diagonal term of our mGNM matrix can also be expressed as
  \begin{eqnarray}\label{eq:2ker_gnm2}
  \Gamma_{ii}= \sum_{j,j\neq i} \frac{ \{w^{1}_{j} \Phi^{1}( r_{ij};\eta^{1} )\}\mu^{2}_i }{a^1\mu^{2}_i+a^2\mu^{1}_i +b\mu^{1}_i\mu^{2}_i},\forall i=1,2,\cdots, N.
 \end{eqnarray}
In this way, the full  multiscale Kirchhoff matrix can be expressed as
 \begin{eqnarray}\label{eq:2ker_gnm3}
 \Gamma_{ij}  = \begin{cases}\begin{array}{ll}
         -\frac{ \{w^{1}_{j} \Phi^{1}( r_{ij};\eta^{1} )\}\mu^{2}_i }{a^1\mu^{2}_i+a^2\mu^{1}_i +b\mu^{1}_i\mu^{2}_i}, &i\neq j \\
         -\sum_{j, j\neq i}^N\Gamma_{ij},  & i=j
							\end{array}
       \end{cases}.
 \end{eqnarray}
 The problem with the matrix in Eq. (\ref{eq:2ker_gnm3}) is that the resulting multiscale Kirchhoff matrix is not symmetric, which may lead to computational difficulty.  To avoid non-symmetric matrix, we further propose an alternative construction to preserve the symmetry of the matrix.

Our basic idea is to determine the diagonal terms $\Gamma_{ii}$ from  Eq. (\ref{eq:relation2}) and then on each row, equally distribute the diagonal term into the non-diagonal parts, under condition that the resulting matrix remains symmetric. To this end, we propose  an iterative scheme as shown in Algorithm \ref{Alg}.

\begin{algorithm}
\caption{Type-2 mGNM multiscale Kirchhoff matrix}\label{mGNM}
\begin{algorithmic}
 \State \textbf{Input: } $ \Gamma_{ii},i=1,2,\cdots,N$  \Comment  Diagonal terms are calculated from mFRI
\vspace{3 mm}
    \For{ $j\gets 2,N$ }   \Comment For the first row and first line of multiscale Kirchhoff matrix.
          \State  $ \Gamma_{1j}=\frac{\Gamma_{11}}{N-1}$   \Comment We equally distribute the diagonal terms into non-diagonal parts.
          \State  $ \Gamma_{j1}=\Gamma_{1j}$               \Comment Use the symmetry property.
     \EndFor
\vspace{3 mm}
  \For{ $i\gets 2,N-1$ }
          \State $sum=0$
      \For{ $k\gets 1,i-1$ }
            \State $k_1=k$
            \State $k_2=k+1$
			\State $sum=sum+\Gamma_{k_1k_2}$              \Comment Summarize over terms already determined from previous iterations.
      \EndFor

     \For{ $j\gets i+1,N$ }
          \State  $ \Gamma_{ij}=\frac{\Gamma_{ii}-sum}{N-i}$   \Comment We equally distribute the diagonal terms into non-diagonal parts.
          \State   $ \Gamma_{ji}=\Gamma_{ij}$                 \Comment Use the symmetry property.
     \EndFor
  \EndFor
 \end{algorithmic}
\label{Alg}
  \end{algorithm}

It also should be noticed that, in the construction of our Type-2 mGNM, only the diagonal terms are fixed and determined from the mFRI. In B-factor prediction, the non-diagonal values can be very flexible as long as they satisfy the network constraint that the summation of their values equals to the diagonal term. We believe this is due to the fact that the success of mGNM in B-factor prediction is determined mostly by the packing information stored in the diagonal terms of its Kirchhoff matrix. In the following discussion, we only use the symmetric scheme in Algorithm \ref{Alg} as our Type-2 mGNM.

\subsection{Multiscale anisotropic network model (mANM)}\label{sec:mANM}

In our mANM, the generalized local $3\times3$ Hessian matrix $H^n_{ij}$ associated with the $n$th kernel  can be written as
\begin{eqnarray}\label{eq:multi-kirchoff1}
 H^{n}_{ij} = -\frac{\Phi^{n}( r_{ij};\eta^{n})}{r_{ij}^2}\left[ \begin{array}{ccc}
	        (x_j-x_i)(x_j-x_i) &(x_j-x_i)(y_j-y_i) &(x_j-x_i)(z_j-z_i)\\
            (y_j-y_i)(x_j-x_i) &(y_j-y_i)(y_j-y_i) &(y_j-y_i)(z_j-z_i)\\
            (z_j-z_i)(x_j-x_i) &(z_j-z_i)(y_j-y_i) &(z_j-z_i)(z_j-z_i)
	      \end{array}\right]  ~ \forall ~ i \neq j.
 \end{eqnarray}
Note that Hinsen \cite{Hinsen:1998} has proposed a special case: $\Phi^{n}( r_{ij};\eta^{n} )= e^{-\left(\frac{r_{ij}}{\eta^n}\right)^2}$.
We further take the diagonal parts as $ H^n_{ii}=-\sum_{i\neq j}H^n_{ij}, \forall i=1,2,\cdots, N$. Basically, it is the summation of all the non-diagonal local matrices.

The key component of our mANM is to construct a multiscale Hessian matrix. Essentially, we employ several Hessian matrices parameterized at different scales and determine their coefficients in the final multiscale Hessian matrix by using our mFRI. It should be noticed that for B-factor prediction, each 3 diagonal terms from the inverse Hessian matrix are summarized together. Therefore, in our Hessian matrix based mFRI, our rigidity index associated with the $n$th kernel is constructed as the summation of the diagonal terms,
 \begin{eqnarray}\label{eq:multi-kirchoff12}
 \mu^n_i= \sum_{i\neq j} \frac{\Phi^{n}( r_{ij};\eta^{n})}{r_{ij}^2}[(x_j-x_i)^2+(y_j-y_i)^2+ (z_j-z_i)^2]=\sum_{i \neq j} \Phi^{n}( r_{ij};\eta^{n}),\forall i=1,2,\cdots, N.
 \end{eqnarray}
It is seen that the rigidity index of mANM defined above is the same as our mFRI rigidity index. Therefore, as far as B-factor prediction is concerned, the mFRI approach for constructing mGNM should work for constructing mANM as well.

We adopt the approach used in Type-1 mGNM to construct  mANM. We propose a multiscale Hessian matrix as  $H =\sum_n a^n H^n$, and the coefficients $a^n$ should be evaluated from
\begin{eqnarray}\label{eq:fit_mANM}
{\rm Min}_{a^{n}} \left\{ \sum_i \left| \sum_{n}a^n \mu^{n}_{i} -\frac{1}{B^e_i}\right|^2\right\}.
\end{eqnarray}
Again, different matrices $\{H^n\}$ should be parametrized at different scales.

{ To summarize the multiscale Gaussian network model and multiscale anisotropic network model, we design a flow chart regarding to their basic procedure as demonstrated in Figure \ref{fig:flow}. }

\tikzstyle{block} = [rectangle, draw, thick, fill=white!20,
    text badly centered,  rounded corners, minimum height=3em, node distance=1.7cm]
\tikzstyle{newblock} = [rectangle, draw, thick,fill=white!20,
     text badly centered,rounded corners, minimum height=3em, node distance=1.7cm]
\tikzstyle{line} = [draw, -latex']
\tikzstyle{cloud} = [draw, ellipse,fill=white!20, thick,node distance=1.7cm,
    minimum height=2em]
\tikzstyle{elliptic} = [rectangle, draw, thick, fill=white!20,
     rounded corners, minimum height=2em, node distance=1.7cm]

\begin{figure}
\begin{center}
\begin{tikzpicture}[node distance = 0.5cm]
\tikzstyle{every node}=[font=\large]
    \node[block](init){Read in pdb data};
    \node[newblock, below of=init](selection){Select kernel functions $\phi^n$ and characteristic distances $\eta^n$};
    \node[block, below of=selection](fri){Calculate rigidity index $\mu^n_i$ (or flexibility index $f_i^n$)};
    \node[block, below of=fri](minimization){Evaluate fitting coefficient {$a^n$}(and b) through the minimization process};
    \node[block, below of=minimization](matrix){Construct multiscale Kirchhoff matrix $\Gamma$ (or multiscale Hessian matrix $H$)};
   \node[block, below of=matrix](eigenvalue){Eigenvalue decomposition};
   \node[block, below of=eigenvalue](results){B-factor evaluation or collective mode analysis};
    \path [line] (init) -- (selection);
    \path [line] (selection) -- (fri);
    \path [line] (fri) -- (minimization);
    \path [line] (minimization) -- (matrix);
     \path [line] (matrix) -- (eigenvalue);
     \path [line] (eigenvalue) -- (results);
\end{tikzpicture}
 \caption{{Work flow of basic procedure in mGNM and mANM.}
 }
 \label{fig:flow}
 \end{center}
 \end{figure}
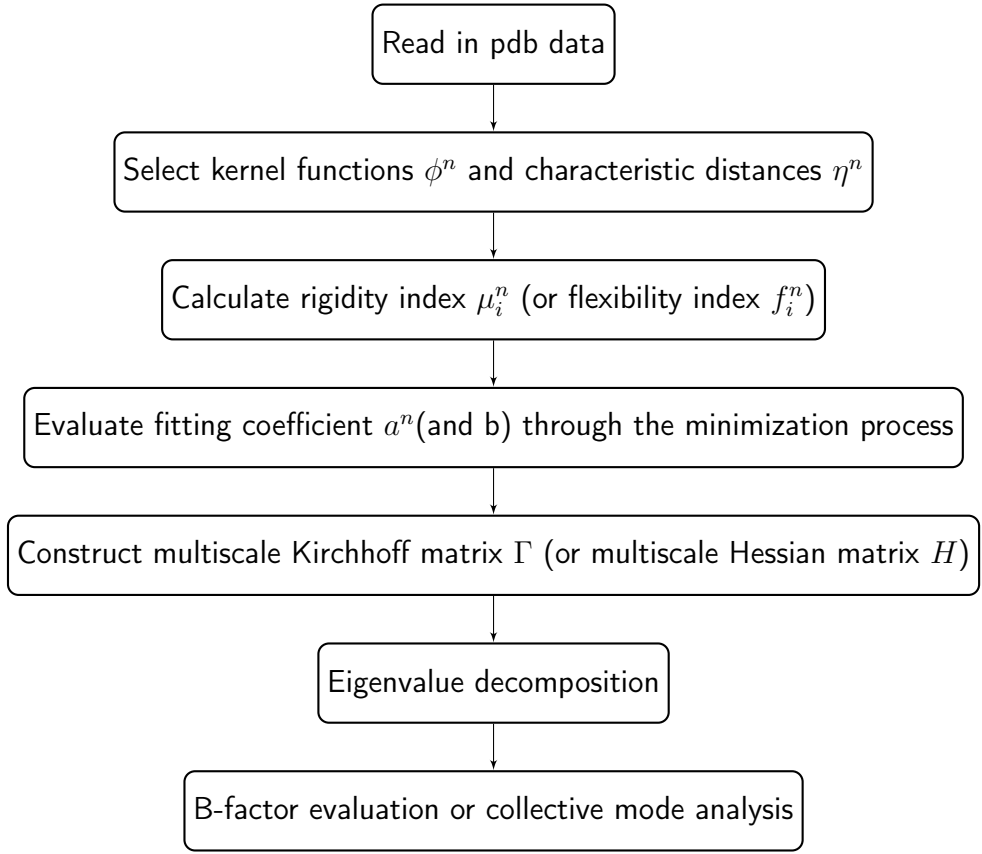

 \section{Validations}\label{sec:result}

 \subsection{The performance of  generalized Gaussian network models }
\paragraph{Comparison between gGNM and FRI}
Based on the analysis in Section \ref{sec:GNM}, it is straightforward to construct correlation function based gGNMs via the matrix inverse of the generalized Kirchhoff matrix (\ref{eqn:MKirchhoff}), which leads to infinitely many new gGNMs, including the original GNM as a special limiting case. It is also possible to construct the FRI by using the Kirchhoff matrix, which gives rise to a unique FRI. Questions arise as what are the relative performance of these correlation function based gGNM and FRI methods. Another question is  whether there is any further relation between these  two distinguished approaches. Specifically, what is the relation between the diagonal  elements of the gGNM matrix inverse  and the FRI direct inverse of the diagonal  elements, for a given generalized Kirchhoff matrix? To answer these questions, we select two representative correlation functions, i.e., the Lorentz ($\upsilon=3$) and ILF functions to construct the generalized Kirchhoff matrix (\ref{eqn:MKirchhoff}). The Lorentz function is a typical example for many   correlation functions studied in our earlier work  \cite{Opron:2014}. In contrast, the ILF function is an extreme case of FRI correlation functions. The resulting two generalized Kirchhoff matrices (\ref{eqn:MKirchhoff}) can be used for calculating the gGNM matrix inverse or the inverse diagonal elements of the FRI matrix. This results in possible combinations or methods, namely, FRI-Lorentz, FRI-ILF, GNM-Lorentz and GNM-ILF. Performances of these methods are carefully analyzed.

\begin{figure}[]
\begin{center}
\begin{tabular}{c}
\includegraphics[width=0.6\textwidth]{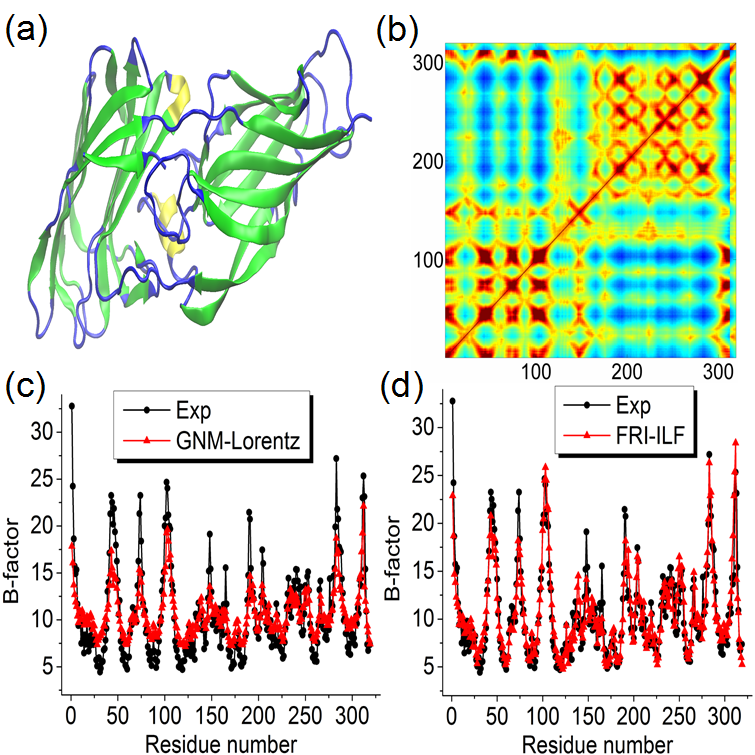}
\end{tabular}
\end{center}
\caption{Illustration of protein 2Y7L.
(a) Structure of protein 2Y7L having two domains;
(b) Correlation map generated by using GNM-Lorentz indicating two   domains;
(c) Comparison of experimental B-factors and those predicted by GNM-Lorentz ($\eta=16$\AA);
(d) Comparison of experimental B-factors and those predicted by FRI-ILF ($r_c=24$\AA).   
 }
\label{2Y7L}
\end{figure}

To answer the above mentioned questions, we first employ a protein from pathogenic fungus Candida albicans (Protein Data Bank ID: 2Y7L) with 319 residues as shown in Fig. \ref{2Y7L}(a) to explore the aforementioned four methods. We consider the coarse-grained  C$_{\alpha}$ representation of protein 2Y7L. We denote $B^{\rm GNM-ILF}$, $B^{\rm FRI-ILF}$, $B^{\rm GNM-Lorentz}$ and $B^{\rm FRI-Lorentz}$ respectively  the predicted B-factors of
GNM-ILF, FRI-ILF, GNM-Lorentz and  FRI-Lorentz methods. The experimental B-factors from X-ray  diffraction,  $B^{\rm Exp}$, are employed for a comparison.  The Pearson product-moment correlation coefficient (PCC) is used to measure  the strength of the linear relationship or dependence between each two sets of B-factors. To evaluate the performance of four methods, we compute the PCCs between predicted B-factors and experimental B-factors. Since performance of these methods depends on their  parameters, i.e.,  the cutoff distance ($r_c$) in the ILF or the scale value ($\eta$) in the Lorentz function, the theoretical B-factors are computed over a wide range of $r_c$ and $\eta$ values.

\begin{figure}[]
\begin{center}
\begin{tabular}{c}
\includegraphics[width=0.8\textwidth]{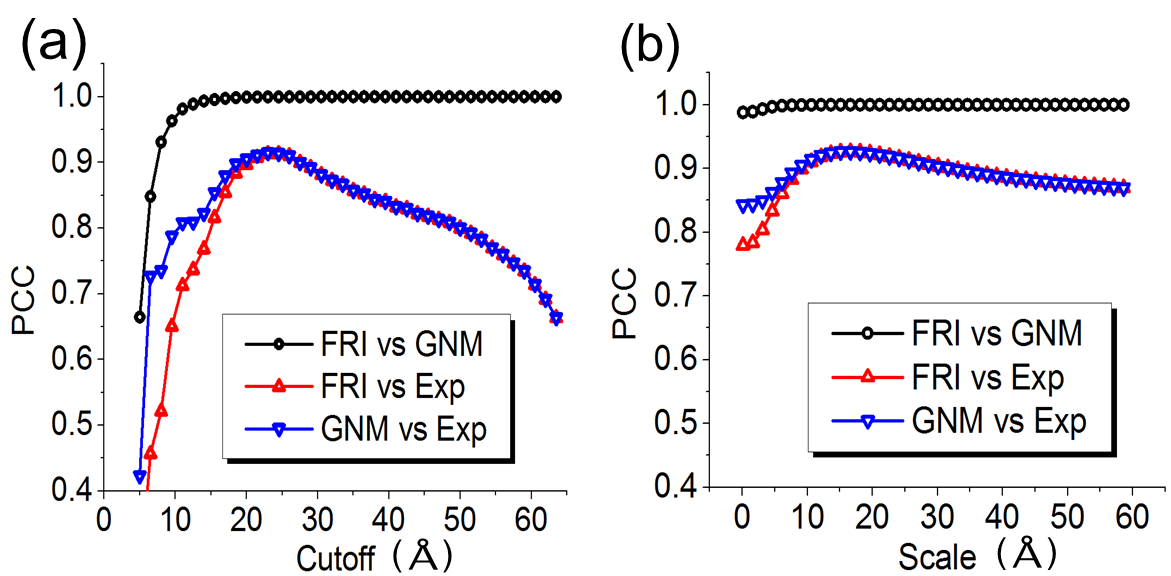}
\end{tabular}
\end{center}
\caption{PCCs  between various B-factors for protein 2Y7L.
(a) Correlations       between $B^{\rm GNM-ILF}$ and $B^{\rm Exp}$,
                       between $B^{\rm FRI-ILF}$ and $B^{\rm Exp}$, and
											 between $B^{\rm GNM-ILF}$ and $B^{\rm FRI-ILF}$;
(b) Correlations       between $B^{\rm GNM-Lorentz}$  and $B^{\rm Exp}$,
                       between $B^{\rm FRI-Lorentz}$  and $B^{\rm Exp}$, and
                       between $B^{\rm GNM-Lorentz}$ and $B^{\rm FRI-Lorentz}$.
}
\label{ccGNMFRI2Y7L}
\end{figure}

Figure \ref{ccGNMFRI2Y7L} depicts PCCs between various B-factors for protein 2Y7L. As shown in  Fig. \ref{ccGNMFRI2Y7L} (a),   the cutoff distance $r_c$ of the ILF is varied from 5\AA~ to 64\AA. The PCCs between $B^{\rm GNM-ILF}$ and $B^{\rm Exp}$, and between $B^{\rm FRI-ILF}$ and $B^{\rm Exp}$, indicate that both  GNM-ILF and    FRI-ILF are able to provide accurate predictions of the experimental B-factors. Their best predictions are attained around $r_c=24$\AA, which is significantly larger than the commonly used GNM cutoff distance of 7-9\AA. 

\paragraph{Intrinsic behavior of gGNM at large cutoff distance}

It is interesting to observe that GNM-ILF and FRI-ILF provide essentially identical predictions when the cutoff distance is equal to or larger than 20\AA.  This phenomenon indicates that when the cutoff is sufficiently large, the diagonal elements of the gGNM inverse matrix and the direct inverse of the diagonal elements of the FRI correlation matrix become linearly dependent.  To examine the  relation between GNM-ILF and FRI-ILF,  we compute PCCs  between   $B^{\rm GNM-ILF}$ and $B^{\rm FRI-ILF}$ over the same range of cutoff distances. As shown in  Fig. \ref{ccGNMFRI2Y7L}(a), there is a strong linear dependence between  $B^{\rm GNM-ILF}$ and $B^{\rm FRI-ILF}$ for $r_c\geq10$\AA. To understand this dependence at large cutoff distance, we consider an extreme case when the cutoff distance is equal to or even larger than the protein size, so all the particles within the network are fully connected. In this situation, we can analytically calculate $i$th diagonal element of the GNM  inverse matrix
\begin{eqnarray}\label{eqn:proprotion0}
\left(\Gamma^{-1}(\Phi( r_{ij};r_c\rightarrow\infty)) \right)_{ii} = \frac{N-1}{N^2},
\end{eqnarray}
and the FRI inverse of the $i$th diagonal element
\begin{eqnarray}  \label{eqn:proprotion01}
 \frac{1}{\sum_{j,j\neq i}^N  \Phi(r_{ij};r_c\rightarrow\infty)}=  \frac{1}{N-1} .
\end{eqnarray}
These results elucidate the strong asymptotic correlation between $B^{\rm GNM-ILF}$ and $B^{\rm FRI-ILF}$ in Fig. \ref{ccGNMFRI2Y7L}(a). They also explain why predictions of the original GNM and FRI-ILF deteriorate as $r_c$ is sufficiently large because all the predicted B-factors become identical, i.e., either $\frac{N-1}{N^2}$ or $\frac{1}{N-1}$. And two methods deliver very similar results, especially when the total number is very large, as we have
$\frac{\frac{N-1}{N^2}}{\frac{1}{N-1}}\rightarrow 1$ when $N\rightarrow\infty$.


The performance and comparison between GNM-Lorentz and FRI-Lorentz are illustrated in Fig. \ref{ccGNMFRI2Y7L}(b) for the scale value $\eta$ from 0.5\AA~ to 64\AA. First, it is seen that the  GNM-Lorentz is a successful new approach. In fact, it outperforms the original GNM  for the peak PCCs. A comparison of the predicted B-factors and the experimental B-factors is plotted in Figs. \ref{2Y7L}(c) and \ref{2Y7L}(d) for  GNM-Lorentz and  FRI-ILF, respectively. It is seen that $B^{\rm FRI-ILF}$ more closely matches the experimental B-factors than $B^{\rm  GNM-Lorentz}$  does due to the different fitting schemes employed by two methods as shown in Eqs. (\ref{eqn:GNM}) and (\ref{eqn:FRI}), respectively.

As shown in Fig.  \ref{ccGNMFRI2Y7L}(b), the predictions from  GNM-Lorentz and FRI-Lorentz become identical as $\eta\geq5$\AA. A strong correlation between $B^{\rm GNM-Lorentz}$ and $B^{\rm FRI-Lorentz}$ is revealed at an even smaller  scale value.  This behavior leads us to speculate a general relation
\begin{eqnarray}\label{eqn:proprotion}
\left(\Gamma^{-1}(\Phi( r_{ij};\eta)) \right)_{ii}
 \longrightarrow
\frac{c}{\sum_{j,j\neq i}^N  \Phi(r_{ij};\eta)},  ~\eta \rightarrow \infty,
\end{eqnarray}
 where $c$ is a constant. Relation (\ref{eqn:proprotion}) means that the correlation function based gGNM  is equivalent to the FRI for a given admissible correlation function when the scale parameter is sufficiently large. This relation is certainly true for the ILF as analytically proved in Eqs. (\ref{eqn:proprotion0}) and (\ref{eqn:proprotion01}).  Relation (\ref{eqn:proprotion})  is  a very interesting and powerful  result not only for sake of understanding GNM and FRI methods, but also for the design  of  accurate and efficient new methods.

It should be noticed that our findings are consistent with the previous finding \cite{Rader:2006} that, the local packing density described by the direct inverse of the diagonal terms represents only the leading order but not the entire set of the dynamics described by  gGNM.  Our results reveal an interesting connection between FRI and gGNM when the characteristic distance is sufficiently large.

\paragraph{Validation of gGNM with extensive experimental data}
\begin{figure}[]
\begin{center}
\begin{tabular}{c}
\includegraphics[width=0.8\textwidth]{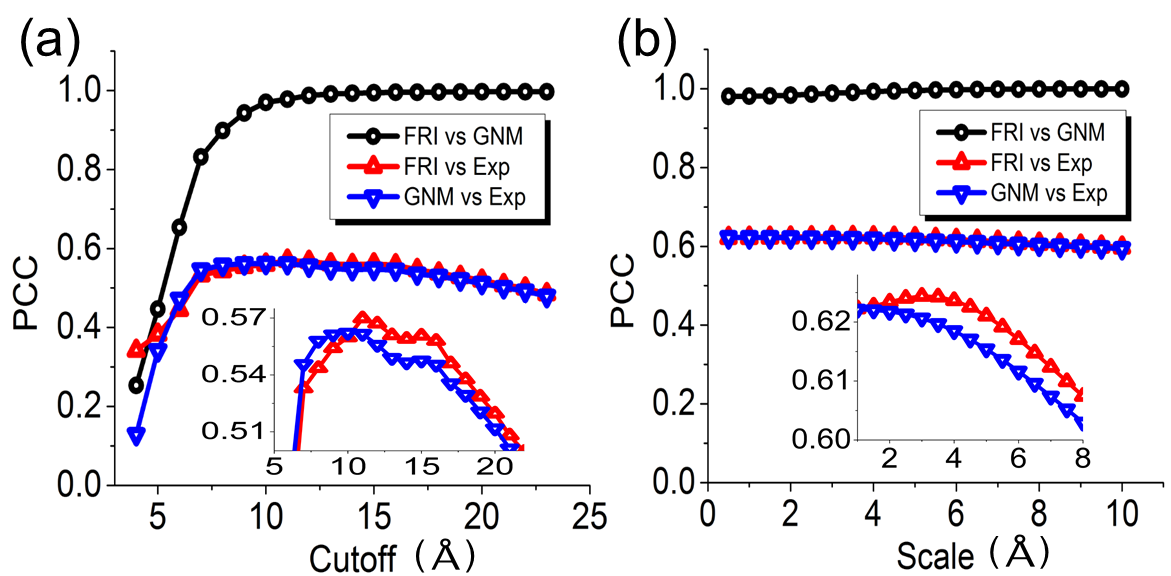}
\end{tabular}
\end{center}
\caption{ PCCs between various B-factors averaged over  364 proteins.
(a) Correlations       between $B^{\rm GNM-ILF}$ and $B^{\rm Exp}$,
                       between $B^{\rm FRI-ILF}$ and $B^{\rm Exp}$, and
											 between $B^{\rm GNM-ILF}$ and $B^{\rm FRI-ILF}$;
(b) Correlations       between $B^{\rm GNM-Lorentz}$  and $B^{\rm Exp}$,
                       between $B^{\rm FRI-Lorentz}$  and $B^{\rm Exp}$, and
                       between $B^{\rm GNM-Lorentz}$ and $B^{\rm FRI-Lorentz}$.
 }
\label{CC_F_G364}
\end{figure}

It remains to prove that the above findings from a single protein are translatable and verifiable to a large class of biomolecules. To this end, we consider a set of 364 proteins, which is a subset of the 365 proteins utilized and documented in our earlier work \cite{Opron:2014}. The omitted protein is 1AGN, which has been found to have unrealistic experimental B-factors.  We carry out systematic studies of four methods over a range of cutoff distances or scale values. For each given $r_c$ or $\eta$, the PCCs between two sets of B-factors are averaged over 364 proteins. Figure \ref{CC_F_G364} illustrates our results. Figure \ref{CC_F_G364}(a) plots the results of the ILF   implemented in both GNM and FRI methods with the cutoff distance varied from 4\AA~ to 23\AA. Figure \ref{CC_F_G364}(b)  depicts  similar results obtained by using the Lorentz function implemented in two methods. The scale value is explored over the range of  0.5\AA~ to 10\AA.
We summarize these results from several aspects as following.

First,  the proposed new method, GNM-Lorentz, is very accurate for the B-factor prediction of 364 proteins as shown in Fig. \ref{CC_F_G364}(b). The best GNM-Lorentz  prediction  is about 10.7\% better than that  of the original GNM shown in Fig. \ref{CC_F_G364}(a). In fact, GNM-Lorentz outperforms the original GNM over a wide range of parameters for this set of proteins, which indicates that  the proposed generalization is practically valuable. Similarly, FRI-Lorentz  is also about 10\% more accurate than FRI-ILF in the B-factor prediction. Since the ILF is a special case and there are  infinitely many FRI correlation functions, there is a wide variety of correlation function based gGNMs that are expected to deliver more accurate flexibility analysis than the original GNM does.

Additionally, the FRI-Lorentz method is able to attain the best average prediction for 364 proteins among four methods as shown in the zoomed in parts in Fig. \ref{CC_F_G364}(b).  However, for a given correlation function, the difference between FRI and gGNM predictions is very small.

Moreover, for a given admissible FRI function, gGNM and FRI B-factor predictions are strongly linearly correlated  and reach near 100\% correlation when $r_c>9$\AA~ or $\eta>0.5$\AA~ for 364 proteins as demonstrated in Fig. \ref{CC_F_G364}.  This finding offers a solid confirmation of  Eq. (\ref{eqn:proprotion}). Therefore, correlation function based gGNMs, including the original GNM as a special case,  are indeed equivalent to the corresponding FRI methods in the flexibility analysis for a wide range of commonly used  scale values.

Furthermore, it has been shown that the fast FRI is a linear scaling method \cite{Opron:2014}, while gGNM scales as $O(N^3)$ due to their matrix inverse procedure. As a result, the accumulated  CPU times for the B-factor predictions of  364 proteins at $r_c=7$ or $\eta=3$ are 0.88, 1.57, 5071.32  and 4934.79 seconds respectively for the FRI-ILF,  FRI-Lorentz, GNM-ILF and  GNM-Lorentz. The test is performed on a cluster with 8 Intel Xeon 2.50GHz CPUs and 128GB memory. In fact, gGNM methods are very fast for small proteins as well.  Most of the accumulated gGNM CPU times are due to the computation of three largest proteins (i.e., 1F8R, 1H6V and 1QKI) in the test set.

Finally, it is worth mentioning that  that the earlier FRI rigidity index includes the contribution from the self correlation, i.e., the diagonal term \cite{KLXia:2013d,Opron:2014}. The present findings do not change if  the summation in the generalized  Kirchhoff matrix (\ref{eqn:MKirchhoff}) is modified to include the diagonal term and then the calculation of gGNM matrix inverse is modified to include the contribution from first eigen mode, i.e., $\left(\Gamma^{-1} \right)_{ii}=\sum_{k=1}^N  \lambda^{-1}\left[{\bf u}_k {\bf u}_k^T \right]_{ii}$.  In fact, this modification makes the   generalized  Kirchhoff matrix less singular and  faster converging.

\subsection{The performance of   multiscale Gaussian network models}\label{sec:multi_gnm}

\paragraph{Type-1 mGNM}
\begin{figure}
\begin{center}
\begin{tabular}{c}
\includegraphics[width=0.9\textwidth]{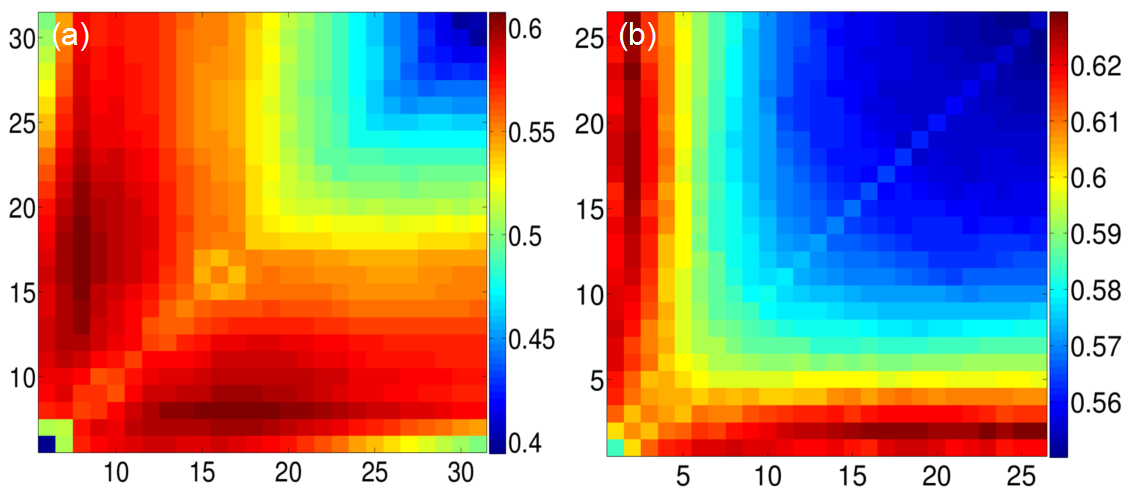}
\end{tabular}
\end{center}
\caption{The average PCCs over 362 proteins for Type-1 mGNM.
(a) Two ILF kernels and their cutoff distances are systematically changed from 5 \AA~ to 31 \AA.
(b) Two exponential kernels and their scales $\eta$ are systematically varied in the range of [1\AA, 26\AA].
}
\label{fig:t1_mGNM_pcc}
\end{figure}

We validate our two types of mGNM with various parameter values over a set of 362 proteins. Two largest proteins, i.e., 1H6V and 1QKI, are removed from our earlier data set of 364 proteins \cite{Opron:2014} due to the limited computational resources. Two kinds of kernels, i.e., ILF and exponential, are employed.  To explore the multiscale behavior,   we use two kernels of the same type but with different characteristic distances in our mGNM schemes. For ILF kernel based test, the cutoff distances in both kernels vary from 5\AA ~ to 31\AA. For exponential kernel based test, we set $\kappa=1$ and vary $\eta$ in both kernels within the range of [1\AA, 26\AA]. The  PCCs with experimental B-factors are averaged over 362 proteins. The results for Type-1 mGNM are demonstrated in Figures \ref{fig:t1_mGNM_pcc} (a) and (b).  When two ILF kernels are used in Figure \ref{fig:t1_mGNM_pcc} (a), we can seen that the largest average PCCs are concentrated around the region where two kernels have dramatically different cutoff distances, i.e., one being around 7 \AA ~ and the other ranging from 14 to 20 \AA.  Our results indicate that   in this set of proteins there is a multiscale property that is better described by mGNM parametrized at different cutoff distances. Moreover, the best PCC is distributed around cutoff distance 7\AA, which  is consistent with the optimal cutoff  distance (7\AA) recommended for  the  traditional GNM method.  Similar multiscale behavior can also be observed for exponential kernel based mGNM as demonstrated in Figure \ref{fig:t1_mGNM_pcc} (b).

\paragraph{Type-2 mGNM}

\begin{figure}
\begin{center}
\begin{tabular}{c}
\includegraphics[width=0.9\textwidth]{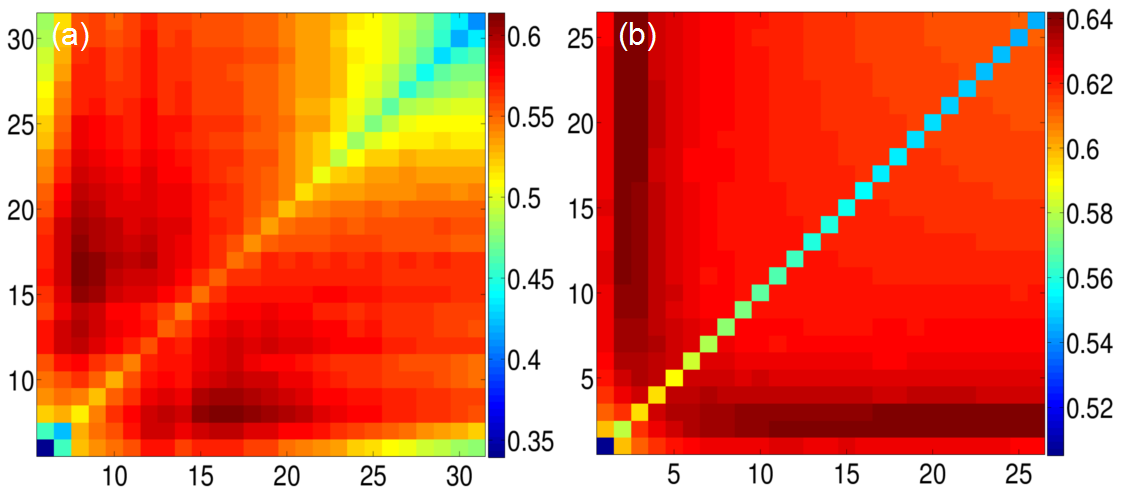}
\end{tabular}
\end{center}
\caption{The average PCCs over 362 proteins for Type-2 mGNM.
(a) Two ILF kernels and their cutoff distances are systematically changed from 5 \AA~ to 31 \AA.
(b) Two exponential kernels and their scales $\eta$ are systematically varied in the range of [1\AA, 26\AA].
}
\label{fig:t2_mGNM_pcc}
\end{figure}

The results of Type-2 mGNM with ILF kernels and exponential kernels are demonstrated in Figures \ref{fig:t2_mGNM_pcc} (a) and (b), respectively. The multiscale property is observed for both cases. Compared with Type-1 mGNM,  Type-2 mGNM is able to achieve better average PCCs with respect to experimental B-factors.  For two ILF kernels, the best average  PCC for traditional GNM is 0.567. Type-1 mGNM  has significantly improved it to 0.607. Additionally, Type-2 mGNM further achieves the best average PCC of 0.614. Similar results are observed in exponential kernel models. For generalized GNM, the best average PCC is about 0.608. This has been enhanced to 0.629 in Type-1 mGNM and further improved to 0.642 in Type-2 mGNM. Detailed comparisons are summarized in Table \ref{tab:four_cases}.

\begin{table}[htbp]
  \centering
	\caption{ The	best average PCCs with experimental B-factors.  Results for  GNM and mGNM are averaged over 362 proteins. Results for ANM and mANM are averaged over 300 proteins.}
 \begin{tabular}{c||c|c|c||c|c|c}
    \toprule
Kernel & GNM & Type-1 mGNM & Type-2 mGNM  & Kernel & ANM & mANM\\
    \midrule
ILF   &0.567   &0.607   &0.614  & ILF  &0.490   &0.531  \\
Exponential &0.608  &0.629   &0.642  &Gaussian   &0.518   &0.546  \\
    \bottomrule
    \end{tabular}%
  \label{tab:four_cases}
\end{table}

\subsection{The performance of   multiscale anisotropic network models}\label{sec:multi_anm}

\begin{figure}
\begin{center}
\begin{tabular}{c}
\includegraphics[width=0.9\textwidth]{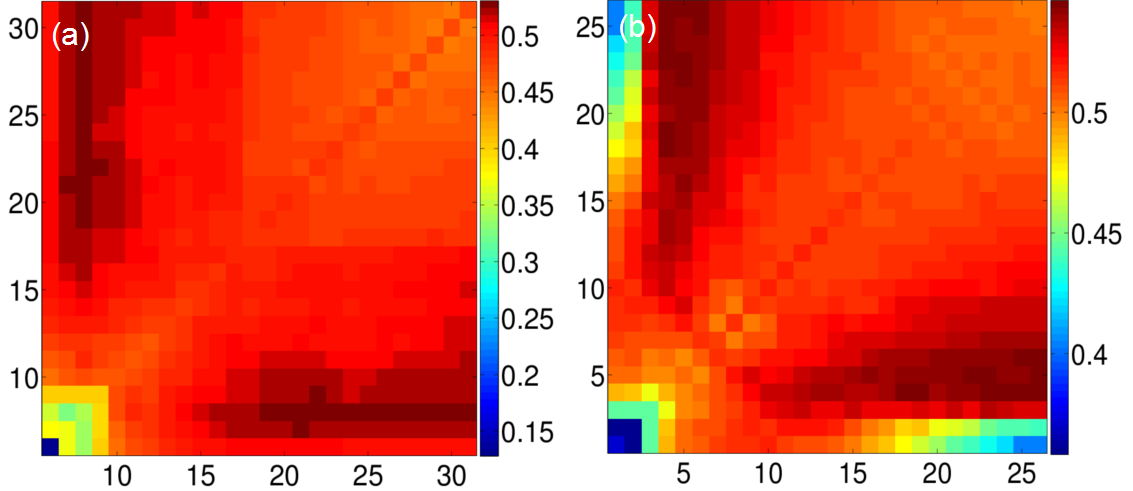}
\end{tabular}
\end{center}
\caption{
The average PCCs over 300 proteins for mANM.
(a) Two ILF kernels and their cutoff distances are systematically changed from 5 \AA~ to 31 \AA.
(b) Two Gaussian kernels ($\kappa=2$) and their scales $\eta$ are systematically varied in the range of [1\AA, 26\AA].
}
\label{fig:mANM}
\end{figure}

\begin{table}[htbp]
  \centering
	\caption{64 Large-sized proteins in the 364-protein data set \cite{Opron:2014} but not included in our mANM test due to limited computational resource.}
 \begin{tabular}{cccccccccc}
    \toprule
1F8R &1GCO  &1H6V  &1IDP  &1KMM  &1QKI  &1WLY  &2A50  &2AH1  &2BCM \\
2COV &2D5W  &2DPL  &2E10  &2ETX  &2FN9  &2I49  &2O6X  &2OKT  &2POF\\
2PSF &2Q52  &2VE8  &2W1V  &2W2A  &2XHF  &2Y7L  &2YLB  &2YNY  &2ZCM \\
2ZU1 &3AMC  &3BA1  &3DRF  &3DWV  &3G1S  &3HHP  &3LG3  &3MGN  &3MRE\\
3N11 &3NPV  &3PID  &3PTL  &3PVE  &3PZ9  &3SRS  &3SZH  &3TDN  &3UR8\\
3W4Q &4AM1  &4B6G  &4B9G  &4DD5  &4DKN  &4DQ7  &4ERY  &4F01  &4G5X\\
4G6C &4J11  &4J78  &4JYP  \\
    \bottomrule
    \end{tabular}%
  \label{tab:mANM_large_protein}
\end{table}
To study the performance of the multiscale anisotropic network model, we use 300 proteins obtained from the dataset with 364 proteins by removing the largest 64 proteins listed in Table \ref{tab:mANM_large_protein}. It should be noticed that the Hessian matrix used in mANM are $3N\times3N$, which is 9 times larger than the correspondent Kirchhoff matrix in gGNM. This poses more challenges as the computational time grows exponentially with the size of the Hessian matrix.

We consider ILF kernel and Gaussian kernel  ($\kappa=2$) based mANM methods in our test study. Our results are plotted in Figure \ref{fig:mANM}. First of all, one can still see the multiscale effect in this set of proteins. The best average PCC values of mANM are achieved at the combination of a relatively small cutoff distance (7\AA) and a relatively large cutoff distance. These values are much higher than those on the diagonal, which are the average PCC values of the traditional (single kernel) ANM.  For Gaussian kernel based mANM, we see a similar pattern. However, it achieves better predictions than those of the ILF kernel based mANM.    This results are also listed in Table   \ref{tab:four_cases}.  Although the ANM methods are not as accurate as the GNM methods, they are able to offer unique collective motions that otherwise cannot be obtained by the GNM methods.

\section{Applications }\label{sec:application}

Having demonstrated the ability of mGNM and mANM for capturing protein multiscale behavior and improving B-factor predictions,  we consider a few applications to showcase the proposed methods.
First, we take on a set of proteins that fail the original GNM in various ways. This analysis might shed light on why the proposed mGNM works better than the original GNM. Additionally,  GNM and ANM can provide domain  information of a protein structure. It is well known that GNM eigenvectors can be used to indicate the possible divisions of domains and domain-domain interactions. Finally,  ANM eigenvectors are widely used to predict the collective motions of a protein near its equilibrium. These issues are investigated in this section.

\subsection{B-factor prediction of difficult cases using mGNM}

\begin{figure}
\begin{center}
\begin{tabular}{c}
\includegraphics[width=0.8\textwidth]{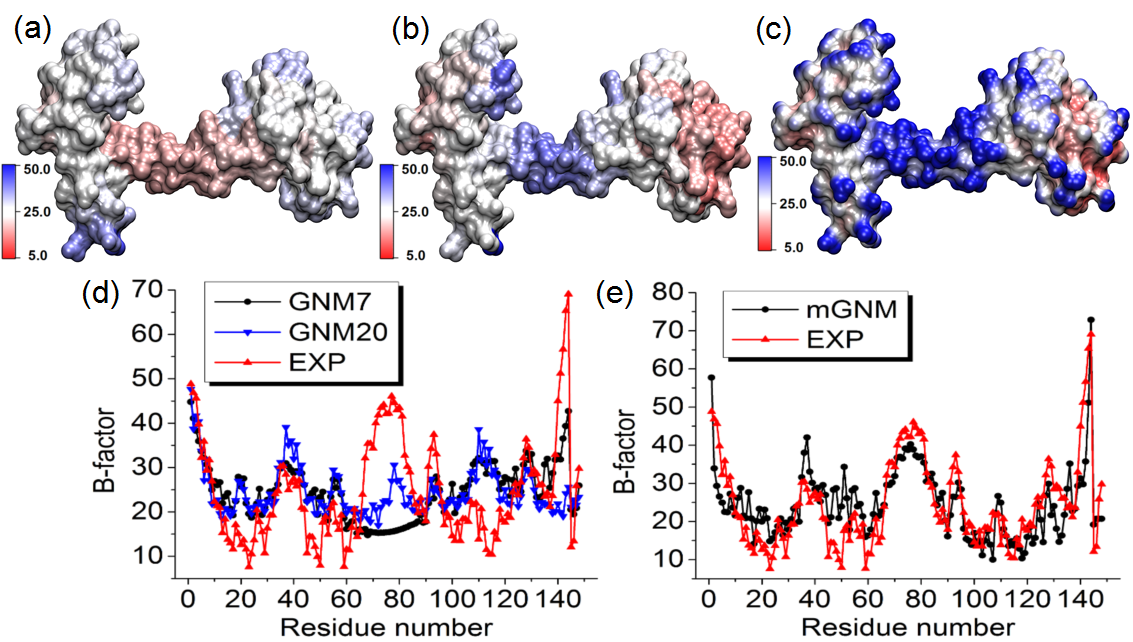}
\end{tabular}
\end{center}
\caption{The comparison between Type-2 mGNM with exponential kernel and traditional GNM for the  B-factor prediction of protein 1CLL. Two scales, i.e.,  $\eta^1=3$\AA~ and $\eta^2=25$\AA, are employed in mGNM. (a) Molecular surface colored by B-factors predicted by GNM with cutoff distance 7 \AA. (b) Molecular surface colored by B-factors evaluated by our Type-2 mGNM.  (c) Molecular surface  colored by multiscale flexibility function in Equation (\ref{eq:mflexibility_function}). (d) B-factors predicted by traditional GNM with cutoff distances 7\AA~ (GNM7) and 20\AA ~(GNM20). (e) B-factors predicted by mGNM.}
\label{fig:1ccl}
\end{figure}

\begin{figure}
\begin{center}
\begin{tabular}{c}
\includegraphics[width=0.8\textwidth]{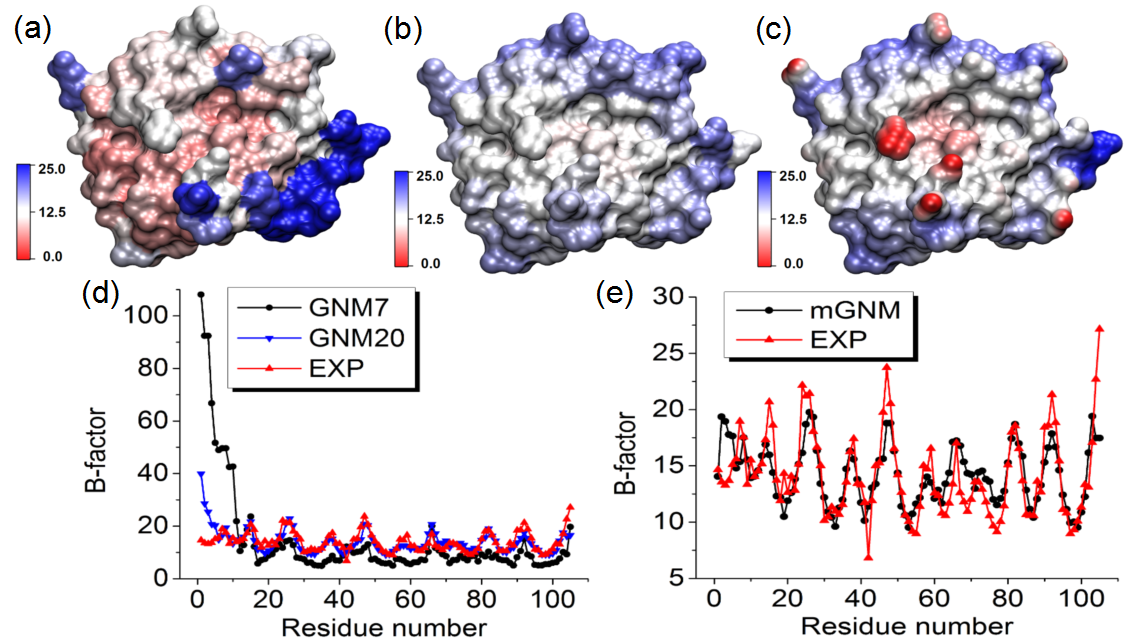}
\end{tabular}
\end{center}
\caption{The comparison between Type-2 mGNM with exponential kernel and traditional GNM for protein 1V70 B-factor prediction. Two scales, i.e.,  $\eta^1=3$\AA~ and $\eta^2=25$\AA, are employed in mGNM. (a) Molecular surface colored by B-factors predicted by GNM with cutoff distance 7 \AA. (b) Molecular surface colored by B-factors evaluated by our Type-2 mGNM.  (c) Molecular surface is colored by multiscale flexibility function in Equation (\ref{eq:mflexibility_function}). (d) B-factors predicted by traditional GNM with cutoff distances 7\AA~ (GNM7) and 20\AA ~(GNM20). (e) B-factors predicted by mGNM. }
\label{fig:1v70}
\end{figure}

\begin{figure}
\begin{center}
\begin{tabular}{c}
\includegraphics[width=0.8\textwidth]{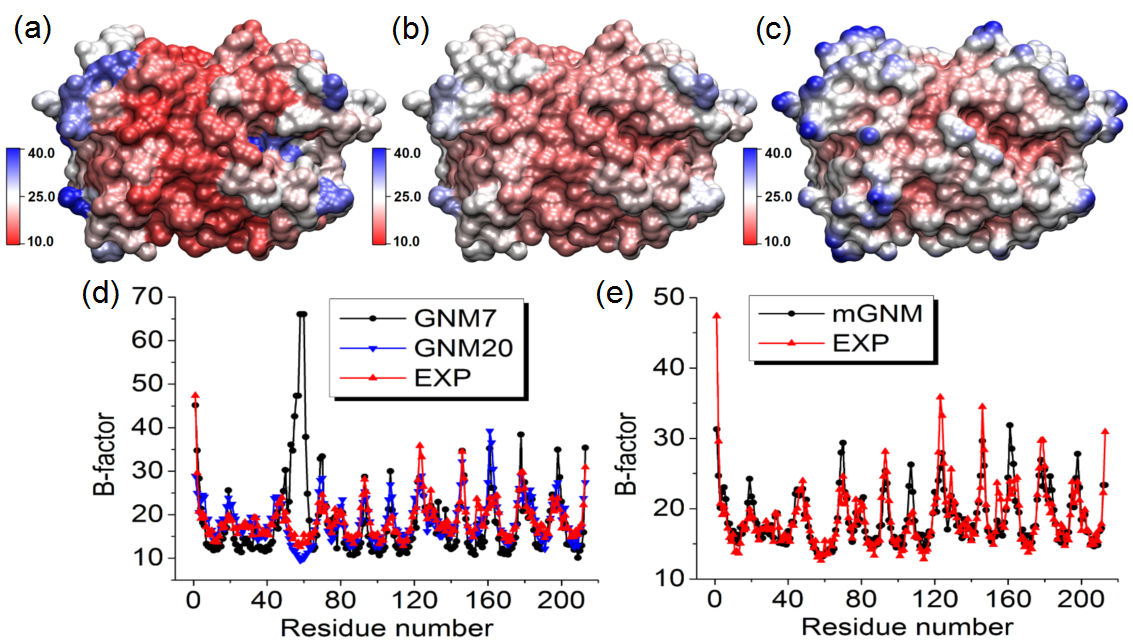}
\end{tabular}
\end{center}
\caption{The comparison between Type-2 mGNM with exponential kernel and traditional GNM for protein 2HQK B-factor prediction. Two scales, i.e.,  $\eta^1=3$\AA~ and $\eta^2=25$\AA, are used in mGNM. (a) Molecular surface colored by B-factors predicted by GNM with cutoff distance 7 \AA. (b) Molecular surface colored by B-factors evaluated by our Type-2 mGNM. (c) Molecular surface is colored by multiscale flexibility function in Equation (\ref{eq:mflexibility_function}).   (d) B-factors predicted by traditional GNM with cutoff distances 7\AA~ (GNM7) and 20\AA ~(GNM20). (e) B-factors predicted by mGNM.}
\label{fig:2hqk}
\end{figure}

\begin{figure}
\begin{center}
\begin{tabular}{c}
\includegraphics[width=0.8\textwidth]{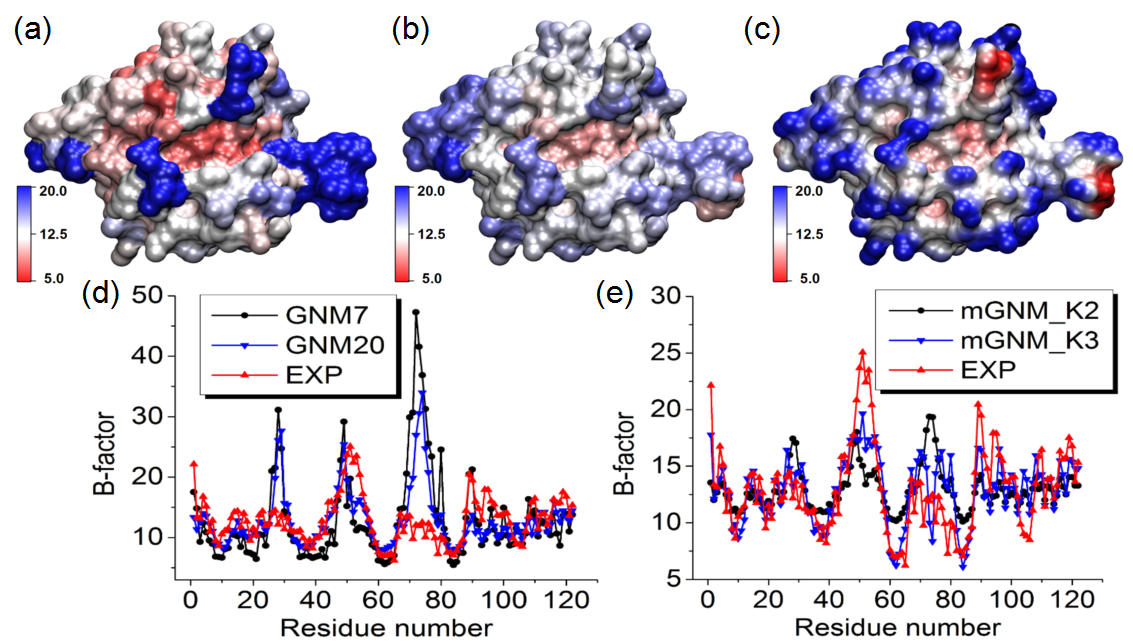}
\end{tabular}
\end{center}
\caption{The comparison between Type-2 mGNM with exponential kernel and traditional GNM for protein 1WHI B-factor prediction. Two mGNMs are used. The first one, mGNM$\_$K2, has two  exponential kernels with $\kappa=1$, $\eta^1=3$\AA~ and $\eta^2=25 $\AA. The second mGNM, mGNM$\_$K3, has an extra  exponential kernel with $\kappa=1$ and $\eta^3=10$ \AA. (a) Molecular surface colored by B-factors predicted by GNM with cutoff distance 7 \AA. (b) Molecular surface colored by B-factors evaluated by our Type-2 mGNM. (c) Molecular surface is colored by multiscale flexibility function in Equation (\ref{eq:mflexibility_function}). (d) B-factors predicted by traditional GNM with cutoff distances 7\AA~ (GNM7) and 20\AA ~(GNM20). (e) B-factors predicted by two mGNMs, i.e.,   mGNM$\_$K2 and  mGNM$\_$K3.}
\label{fig:1whi}
\end{figure}

It is well known that the traditional GNM does not work well in the B-factor prediction for certain proteins for various reasons \cite{JKPark:2013,Opron:2015a}. Park et al. have shown that GNM PCCs with experimental B-factors can be negative \cite{JKPark:2013}. In this work, we demonstrate that the present mGNM is able to deliver good B-factor predictions by capturing multiscale features. To this end,  we consider four proteins, i.e., 1CLL, 1V70, 2HQK and 1WHI. The Type-2 mGNM with two exponential kernels is utilized in our study.  As depicted in Figure \ref{fig:t2_mGNM_pcc}(b), there is a wide range of scale parameters that deliver accurate B-factor predictions. We simply choose $\kappa=1, \eta^1=3$ \AA~ and $\kappa=1, \eta^2=25$\AA~ in our studies. To draw a comparison, the traditional GNM, i.e., GNM-ILF, is employed with different cutoff distances, namely 7 and 20 \AA, which are denoted as  GNM7 and GNM20, respectively.

Figures \ref{fig:1ccl}, \ref{fig:1v70}, \ref{fig:2hqk} and \ref{fig:1whi} illustrate our results. In each figure, protein surfaces are colored by B-factor values predicted by GNM7, mGNM and the flexibility function in Eq. (\ref{eq:mflexibility_function}), respectively in subfigures (a), (b) and (c). The comparisons of B-factors predicted by GNM7 and GNM20 with those of experiments are demonstrated in subfigures (d).  Similarly, the comparisons of the predicted B-factors by mGNM with those of experiments are plotted in subfigures (e). A summary of related PCC values are listed in Table \ref{tab:four_cases}.

\begin{table}[htbp]
  \centering
	\caption{Case study of B-factor prediction for four proteins in three different schemes, i.e., GNM7, GNM20 and mGNM. In the case of 1WHI, we use mGNM with two kernels and three kernels (value in parentheses).  }
 \begin{tabular}{cccc}
    \toprule
PDB ID& GNM7 & GNM20 & mGNM \\
    \midrule
1CLL & 0.261 & 0.235 & 0.763  \\
1V70 & 0.162 & 0.548 & 0.750  \\
2HQK & 0.365 & 0.781 & 0.833  \\
1WHI & 0.270 & 0.370 & 0.484(0.766) \\
    \bottomrule
    \end{tabular}%
  \label{tab:mGNM}
\end{table}

Flexible hinges are important to protein function, but may not be easily detected by GNM type of methods \cite{stonehinge, flexoracle}. As shown in Figure \ref{fig:1ccl}, the original GNM parametrized at cutoff distance 7 or 20 \AA~ does not work well for the hinge located around residues 65-85.   In fact, the GNM method cannot predict the flexible hinge  at any given cutoff distance. Whereas, our two-kernel mGNM is able to capture the hinge behavior.

Protein 1V70 shown in Figure \ref{fig:1v70} is another difficult case for the traditional GNM method. At cutoff distance 7\AA, it severely over-predicts the B-factors of the first 12 residues.  However, its prediction improves if a  larger cutoff distance is used. In contrast, our two-kernel mGNM provides a very good prediction.

Figure \ref{fig:2hqk} illustrates one more interesting situation. The tradition GNM with cutoff distance 7\AA~
over-predicts the B-factors for residues near number 58. However, at a large cutoff distance of 20\AA, it is able to offer accurate results. In this case, our mGNM is able to further improve the accuracy.

The case of 1WHI given in Figure  \ref{fig:1whi} is  difficult. The GNM with two different parametrizations does not work well. However, our two-kernel mGNM does not work well either. Its PCC of 0.484 is just a minor improvement of GNM values 0.270 (obtained at $r_c=7$\AA) and 0.370 (obtained at $r_c=20$\AA). It should be noticed that our mGNM can simultaneously incorporate several scales. Therefore, we employ an extra kernel with $\kappa=1, \eta^3=10$ \AA~ to deal with this protein.  As shown in Table \ref{tab:mGNM} and Figure  \ref{fig:1whi}, our three-kernel mGNM is able to deliver a good PCC of 0.766.

\subsection{Domain decomposition using mGNM}

\begin{figure}
\begin{center}
\begin{tabular}{c}
\includegraphics[width=0.8\textwidth]{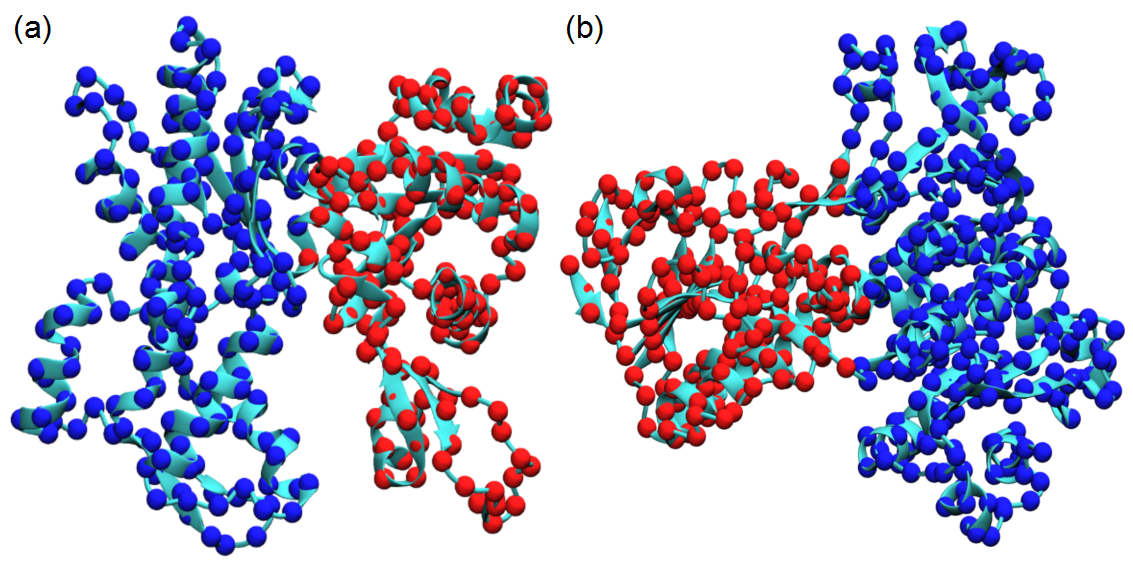}
\end{tabular}
\end{center}
\caption{ Protein domain decomposition with Type-1 mGNM. The first eigenvector (Fiedler vector) is used to decompose the protein into two domains. (a)  protein 1ATN (chain A); (b) protein 3GRS.
 }
\label{fig:mGNM_Type-1}
\end{figure}

Mathematically, the first smallest nonzero eigenvalue is called algebraic connectivity or Fiedler value and the related eigenvector is called Fiedler vector. It is known that the Fiedler vector can be used to decompose a protein into two domains. The way to subdivide a protein is quite natural. As each particle in the protein is assigned with a value (element) from the Fiedler vector, one simply groups these particles according to their positive or negative signs. More specifically, all atoms with positive values are in the same group and the others with negative values are in other group. The ones with zero values can be classified into either group as their are usually the link region between two domains.

\begin{figure}
\begin{center}
\begin{tabular}{c}
\includegraphics[width=0.8\textwidth]{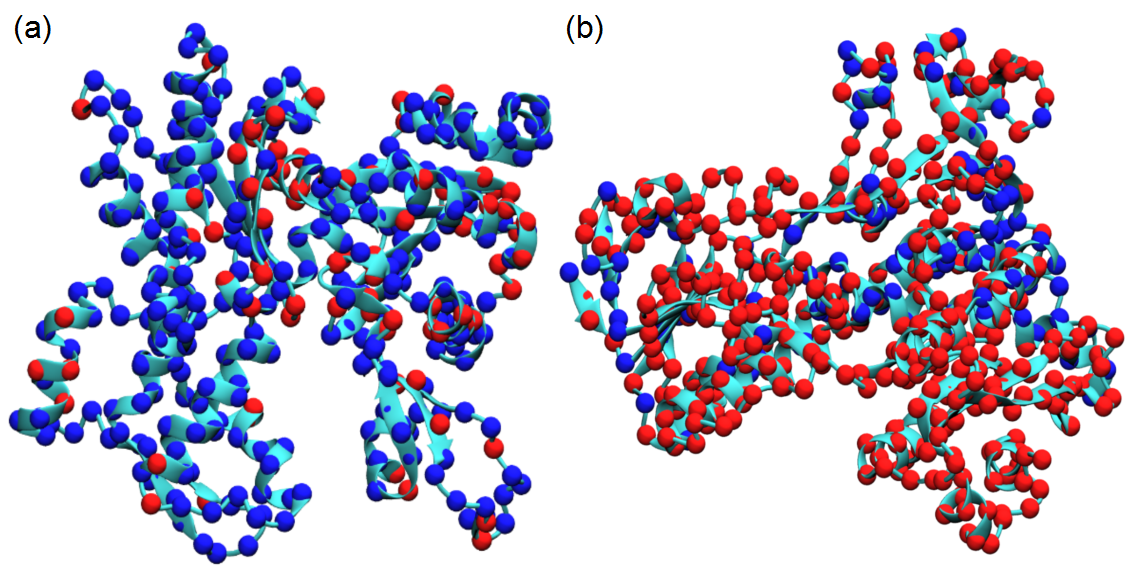}
\end{tabular}
\end{center}
\caption{ Protein domain decomposition with Type-2 mGNM. The first eigenvector (Fiedler vector) is used to decompose the protein into two domains. (a)  protein 1ATN (chain A); (b) protein 3GRS. It can be seen that Type 2 mGNM fails in protein domain decomposition.
 }
\label{fig:mGNM_Type-2}
\end{figure}
To test the performance of  our mGNM schemes,  we adopt two test proteins, i.e., 1ATN (chain A) and 3GRS, used by Kundu, et al. \cite{Kundu:2004}  We compare the performance of our two types of mGNM. In Type-1 mGNM, we use the two exponential kernels  with $\kappa=1, \eta^1=3$ \AA~ and $\kappa=1, \eta^2=25$\AA. In Type-2 mGNM, we use three exponential kernels with an extra kernel parametrized as $\kappa=1, \eta^3=10$ \AA. Our results are depicted in Figures \ref{fig:mGNM_Type-1} and  \ref{fig:mGNM_Type-2}, respectively. It can be seen that Type-1 mGNM delivers a great decomposition, which is also consistent with the prediction from traditional GNM \cite{Kundu:2004}. However, the Type-2 mGNM does not produce a reasonable result. This is due to the fact that
Algorithm \ref{Alg} is designed to construct the symmetric Kirchhoff matrix with required diagonal elements. Its non-diagonal elements do not properly reflect the protein connectivity.

However, we should notice that  the PCCs of Type-1 mGNM for 1ATN and 3GRS are 0.460 and 0.658, respectively. Whereas, the PCCs of Type-2 mGNM for  1ATN and 3GRS are  0.660 and 0.666, respectively. These results indicate that the B-factor values are mainly dictated by the diagonal matrix elements, while the domain separation is determined by non-diagonal matrix elements, which reflect the protein connectivity  in Type-1 mGNM, but have little  to do with the packing geometry in Type-2 mGNM.

\subsection{Collective motion  simulation using mANM}

\begin{figure}
\begin{center}
\begin{tabular}{c}
\includegraphics[width=0.8\textwidth]{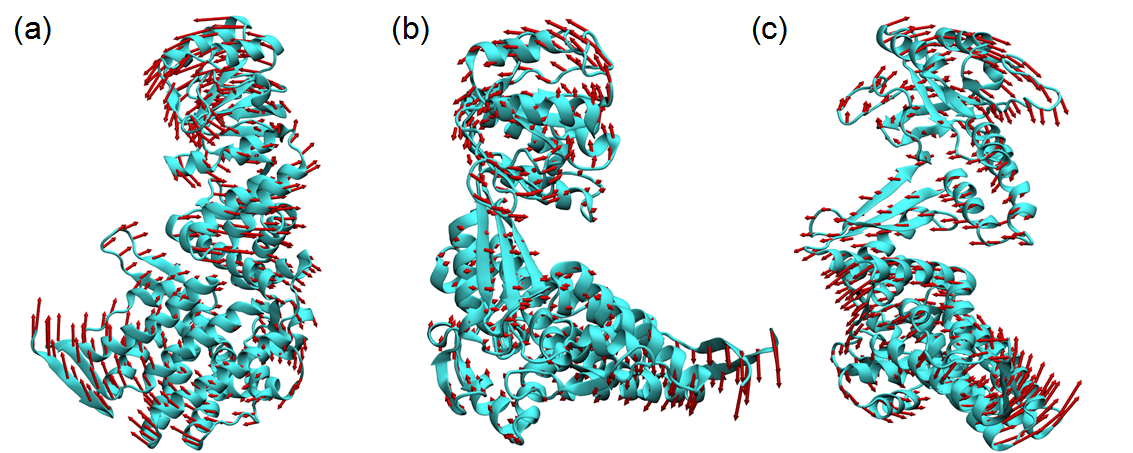}
\end{tabular}
\end{center}
\caption{The collective motions of protein 1GRU (chain A). The seventh, eighth and ninth modes calculated from our mANM are demonstrated in (a), (b) and (c), respectively.}
\label{fig:mANM_1gru}
\end{figure}

\begin{figure}
\begin{center}
\begin{tabular}{c}
\includegraphics[width=0.8\textwidth]{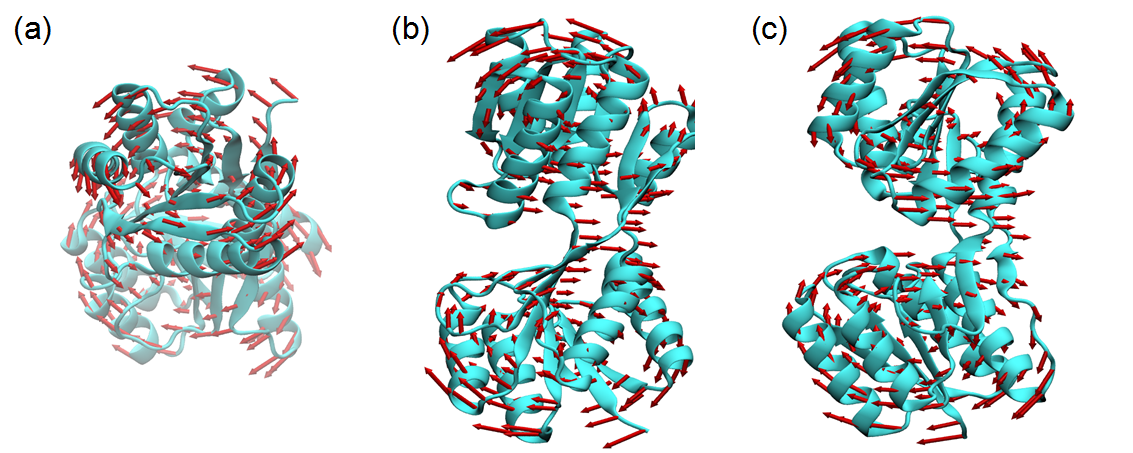}
\end{tabular}
\end{center}
\caption{ The collective motions of protein 1URP (chain A). The seventh, eighth and ninth modes calculated from our mANM are demonstrated in (a), (b) and (c), respectively.}
\label{fig:mANM_1urp}
\end{figure}

GNM is an isotropic model which quantifies the atomic scalar  fluctuations in molecule. In contrast, ANM is designed to describe the anisotropic properties, such as collective motions  of a molecule near the equilibrium.  Typically, the first six modes, corresponding to six zero (or near zero) eigenvalues, represent the trivial translational and rotational modes of a complex biomolecule. Global modes that are unique to the biomolecular structure are described by eigenvectors associated with the nonzero (next smallest) eigenvalues. Due to its simplicity, ANM is widely used to study the dynamics of biomolecules.

In the present work, we have designed our mANM to maintain the aforementioned properties.  To validate our mANM for anisotropic mode analysis, we use two test proteins, i.e., 1GRU (chain A) and 1URP (chain A).  The protein 1GRU is chaperonin GroEL, a benchmark test for ANM \cite{WZheng:2007,Uyar:2014}. We employ our mANM with two Gaussian kernels ($\kappa=2$) with $\eta=5$\AA~ and $\eta=20$\AA. We compute eigenvectors associated with the first three nonzero eigenvalues. As illustrated in Figure \ref{fig:mANM_1gru}, our mANM results  are in an excellent agreement  with  those  of ANM for chaperoin GroEL \cite{WZheng:2007,Uyar:2014}. 

To further validate our mANM, we examine another test case, 1URP. It is   a ribose-binding protein  and its anisotropic motions have been studied in the literature \cite{HYLi:2013}. We utilize the same set of parameters described above. Figure \ref{fig:mANM_1urp} demonstrates mANM results. Our results are in a close  consistence with the traditional ANM analysis \cite{HYLi:2013}. 

\section{Conclusion}\label{sec:concluding}

Gaussian network model (GNM) and anisotropic network model (ANM) are popular methods for macromolecular flexibility analysis. Alternative methods, flexibility-rigidity index (FRI)  \cite{KLXia:2013d,Opron:2014} and anisotropic flexibility-rigidity index (aFRI)  \cite{Opron:2014},  have been introduced to achieve better accuracy and more adaptivity in our recent work. Most recently, we have further proposed multiscale flexibility-rigidity index (mFRI)  \cite{Opron:2015a} to capture the multiscale behavior in macromolecules. Our mFRI utilizes multiple kernels which are parametrized at different scales to describe macromolecular multiscale connectivity. We have shown that mFRI is about 20\% more accurate than GNM in the B-factor prediction of a set of 364 protein \cite{Opron:2015a}. Motivated by these achievements, we propose a few FRI based generalizations of GNM and ANM in this work.

 {First, we construct a series of generalized Gaussian network models (gGNMs). We show that the original Kirchhoff matrix used in GNM can be constructed by using the ideal low-pass filter (ILF),  which is  a special case of a family of admissible correlation kernels (or functions) used in FRI. Based on this connection, we propose a unified framework to construct generalized Kirchhoff matrices for both GNM and FRI. More specifically, the inverse of the generalized Kirchhoff matrices leads to infinitely many gGNMs and the direct inverse of the diagonal terms gives rise to FRI. We  reveal the identical behavior between gGNM and FRI at a large cutoff distance or characteristic scale for B-factor protein predictions.
Additionally, we propose multiscale Gaussian network models (mGNMs)  based on the relationship of GNM and FRI. Essentially, we develop a two-step procedure to construct mGNMs. In the first step, we utilize  mFRI to come up with an optimal combination of multiscale kernels. In our second, we try to implement the same combination of multiscale kernels in the  generalized Kirchhoff matrices for mGNMs. However, this step is not unique because for a given Kirchhoff matrix, GNM and FRI are connected only through diagonal elements. Two types schemes,  Type-1 mGNM and Type-2 mGNM, are proposed in this work.
Moreover, we propose multiscale anisotropic network models (mANMs) based on the similarity between  ANM and GNM  and the connection between GNM and FRI.  Since ANM is typically less accurate than GNM in B-factor prediction \cite{JKPark:2013, Opron:2014}, its main utility is for collective motion analysis. We therefore have developed mANMs to maintain the physical connectivity of protein atoms in the  Kirchhoff matrix. }

 { We have carried out intensive numerical experiments to validate the proposed gGNM, mGNM and mANM methods for B-factor predictions. The gGNM method is examined over a set of 364 proteins. It is found that the proposed gGNM is about 10\% more accurate than GNM in B-factor prediction. For mGNM, we use only a set of 362 protein due to limited computer resource. We show that mGNM can achieve  about 13\% improvement over GNM. Similarly, the proposed mANM is about 11\% more accurate than its counterpart, ANM, in B-factor prediction over a set of 300 proteins.
Further, we consider three types of applications of the proposed mGNM and mANM methods. One type of application is to analyze the flexibility of proteins that fail the original GNM method in various ways. We employ four proteins  to demonstrate the advantage  of the proposed mGNM in flexibility analysis.    Another application is the study of protein domain separations. The first nontrivial eigenmode of the multiscale Kirchhoff matrix is used. We found from the analysis of two proteins that Type-1 mGNM does a good job in domain analysis while Type-2 mGNM does not work for this purpose. The other application concerns the protein collective motions.  Our mANM is found to offer similar results as those of the original ANM method.}
{In the future, we will further apply our mANM to study the anisotropic B-factor \cite{Eyal:2007} and conformational change \cite{Tama:2001}. }

It is worth to pointing that our mGNM and mANM methods are not unique. How to design optimal new mGNM and mANM methods is still an open problem. Essentially, we hope these new methods are efficient, accurate and robust. More specifically, high accuracy in B-factor prediction is a main criterion. Additionally, having the ability to provide correct protein domain analysis  is a desirable property as well.  For mANM, the capability of offering correct motion analysis is a major requirement. The quality of both domain and motion analyses depends on how to design non-diagonal matrix elements so as to properly reflect the physical connectivity among particles.

\vspace*{1cm}

\section*{Acknowledgments}

This work was supported in part by NSF grants   IIS-1302285,  and DMS-1160352, and NIH Grant R01GM-090208.
The authors  thank Jianyu  Chen and Michael Feig for useful  discussions.
\vspace*{1cm}

\end{document}